\documentclass[aps,prb,amsmath,amssymb,eqsecnum,twocolumn,showpacs]{revtex4-1}

\usepackage[utf8]{inputenc}

\usepackage{graphicx}
\usepackage{subfigure}
\usepackage[colorlinks=true,citecolor=blue]{hyperref}
\usepackage{hypernat}
\usepackage{color}%only needed for coloured text

\usepackage{verbatim} %mehrzeilige Kommentare

\newcommand{\bra}[1]{\langle #1|}
\newcommand{\ket}[1]{|#1\rangle}
\newcommand{\RPG}[1]{\re\Psi\left(\frac{1}{2}+i\frac{\beta#1}{2\pi}\right)} %real part of polygamma function of 1/2+i\beta/(2pi)*x
\renewcommand{\vec}[1]{\mathbf{#1}}
\DeclareMathOperator{\re}{\text{Re}}
\DeclareMathOperator{\im}{\text{Im}}

\def\up{\uparrow}
\def\down{\downarrow}
\def\kB {k_\text{B}}
\newcommand{\fmL}{f_\text{L}^-}
\newcommand{\fpL}{f_\text{L}^+}
\newcommand{\fmR}{f_\text{R}^-}
\newcommand{\fpR}{f_\text{R}^+}

\begin{document}

\title{Transport through quantum-dot spin valves containing magnetic impurities}

\author{Bj\"orn Sothmann}
\author{J\"urgen K\"onig}
\affiliation{Theoretische Physik, Universit\"at Duisburg-Essen and CeNIDE, 47048 Duisburg, Germany}

\date{\today}

\begin{abstract}
We investigate transport through a single-level quantum dot coupled to noncollinearly magnetized ferromagnets in the presence of localized spins in either the tunnel barrier or on the quantum dot. For a large, anisotropic spin embedded in the tunnel barrier, our main focus is on the impurity excitations and the current-induced switching of the impurity that lead to characteristic features in the current. In particular, we show how the Coulomb interaction on the quantum dot can provide more information from tunnel spectroscopy of the impurity spin. In the case of a small spin on the quantum dot, we find that the frequency-dependent Fano factor can be used to study the nontrivial, coherent dynamics of the spins on the dot due to the interplay between exchange interaction and coupling to external and exchange magnetic fields.
% We investigate transport through a single-level quantum dot coupled to noncollinearly magnetized ferromagnets in the presence of localized spins in either the tunnel barrier or on the quantum dot. For a spin embedded in the tunnel barrier, we find an interplay between current-induced switching of the spin, spin-dependent tunneling through the barrier and spin accumulation on the dot resulting in characteristic signals in the current. We, furthermore, find huge Fano factors due to random telegraph noise. For noncollinear geometries, an exchange field that depends on the impurity spin state leads to characteristic fingerprints in the transport properties. In the case of a spin on the quantum dot, we find that the frequency-dependent Fano factor can be used to study the nontrivial dynamics of the spins on the dot due to the interplay between exchange interaction and coupling to external and exchange magnetic fields.
\end{abstract}

%73.23.Hk 	Coulomb blockade; single-electron tunneling 
%72.25.Mk 	Spin transport through interfaces 
%85.75.-d 	Magnetoelectronics; spintronics: devices exploiting spin polarized transport or integrated magnetic fields
%72.70.+m 	Noise processes and phenomena
%73.63.Kv 	Quantum dots
%71.70.-d 	Level splitting and interactions
%71.70.Gm 	Exchange interactions 
%71.70.Jp 	Nuclear states and interactions 

%\keywords{}

\maketitle

\section{\label{sec:introduction}Introduction}
Recently, there has been growing interest in spin-dependent transport through nanostructures due to possible applications for spintronics devices.
Transport through quantum dots coupled to ferromagnetic electrodes is particularly interesting due to the interplay of strong Coulomb interaction on the quantum dot and nonequilibrium physics. 
Such systems have been realized experimentally, e.g., by coupling self-assembled semiconducting quantum dots,~\cite{hamaya_spin_2007,hamaya_electric-field_2007,hamaya_kondo_2007,hamaya_oscillatory_2008,hamaya_tunneling_2008} small metallic grains,~\cite{deshmukh_using_2002,bernand-mantel_evidence_2006,mitani_current-induced_2008,bernand-mantel_anisotropic_2009,birk_magnetoresistance_2010} quantum dots defined in InAs nanowires,~\cite{hofstetter_ferromagnetic_2010} carbon nanotubes,~\cite{jensen_single-wall_2003,sahoo_electric_2005,jensen_hybrid_2004,hauptmann_electric-field-controlled_2008,merchant_current_2009} or even single C$_{60}$ molecules~\cite{pasupathy_kondo_2004} to ferromagnetic electrodes.
Quantum dots coupled to ferromagnets have been also studied extensively from a theoretical point of view.~\cite{buka_current_2000,weymann_zero-bias_2005,weymann_tunnel_2005,weymann_cotunneling_2006,fransson_angular_2005,fransson_angular_2005-1,
cottet_positive_2004-1,cottet_positive_2004,cottet_magnetoresistance_2006,martinek_kondo_2003,martinek_kondo_2003-1,martinek_gate-controlled_2005,sindel_kondo_2007,simon_kondo_2007}
Especially interesting are quantum-dot spin valves, i.e., a single-level quantum dot coupled to noncollinearly magnetized electrodes.~\cite{kaenig_interaction-driven_2003,braun_theory_2004,utsumi_nonequilibrium_2005,braun_frequency-dependent_2006,weymann_cotunneling_2007,splettstoesser_adiabatic_2008,lindebaum_spin-induced_2009,sothmann_probing_2010}
Here, spin-dependent tunneling leads to a spin accumulation on the dot that in turn influences the transport through the system by blocking the current. In addition, there is an exchange field acting on the dot spin. It is due to quantum charge fluctuations on the quantum dot and relies on the strong Coulomb interaction on the dot. It gives rise to a precession of the accumulated spin, thereby lifting the spin blockade of the dot. The resulting interplay of spin accumulation and spin precession leads to a number of interesting features in the transport characteristics, e.g., a shift of the peaks in the linear conductance with the angle enclosed by the magnetizations,~\cite{kaenig_interaction-driven_2003,braun_theory_2004} a broad region of negative differential conductance in nonlinear transport~\cite{braun_theory_2004} as well as to characteristic features in the finite-frequency Fano factor~\cite{braun_frequency-dependent_2006} and a splitting of the Kondo resonance.~\cite{martinek_kondo_2003,martinek_kondo_2003-1,martinek_gate-controlled_2005,sindel_kondo_2007,simon_kondo_2007}

While in the above studies all the spin dynamics takes place in the singly occupied orbital of the quantum dot, more complex spin dynamics appears when additional spin excitations are possible. These may be spin waves in the ferromagnetic leads, as discussed in Ref.~\onlinecite{sothmann_influence_2010}.
In the present paper, we investigate a different situation, namely, the coupling to a magnetic impurity either with a large, anisotropic spin or with a spin $1/2$.

We consider two different scenarios.
In the first scenario, a magnetic impurity with a large, anisotropic spin is embedded in one of the tunnel barriers of a quantum-dot spin valve. Here, our main focus is on the spectroscopy of the impurity spin as well as on its switching by the spin-polarized current.  We consider the impurity spin in one of the barriers as this leads to a simpler conductance pattern (only processes involving this particular barrier can excite the spin), thereby simplifying the analysis of the impurity spin behavior.
Our system is somewhat related to  the case of transport through single tunnel barriers containing a magnetic atom or a single molecular magnet, that has been investigated extensively in the recent past, both from a theoretical~\cite{kim_electronic_2004,misiorny_quantum_2007,misiorny_magnetic_2007,fernandez-rossier_theory_2009,fransson_spin_2009,persson_theory_2009,lorente_efficient_2009,delgado_spin-transfer_2010,sothmann_nonequilibrium_2010} as well as from an experimental~\cite{heinrich_single-atom_2004,hirjibehedin_spin_2006,meier_revealing_2008,hirjibehedin_large_2007,otte_role_2008,otte_spin_2009,loth_controlling_2010} point of view. It was shown that the steps observed in the differential conductance can be used to extract magnetic properties such as anisotropies of the atomic spin.~\cite{hirjibehedin_large_2007,fernandez-rossier_theory_2009,fransson_spin_2009,persson_theory_2009,lorente_efficient_2009} Furthermore, the influence of nonequilibrium spin occupations was discussed,~\cite{sothmann_nonequilibrium_2010} explaining the overshooting observed at the conductance steps in experiment and predicting a super-Poissonian current noise. The absence of certain nonequilibrium features in turn was interpreted in terms of an anisotropic relaxation channel. For systems with magnetic electrodes, the possibility to switch the embedded spin by the spin-polarized current through the barrier was predicted theoretically~\cite{delgado_spin-transfer_2010,misiorny_quantum_2007,misiorny_magnetic_2007} and observed in experiment.~\cite{loth_controlling_2010}

In the model studied in this paper, the tunnel barrier containing the magnetic impurity connects a lead with a quantum dot.
The transport behavior is, in this case, more complex since the spin dynamics of the embedded impurity is coupled to the charge and spin degrees of freedom of the quantum dot. 
This has several consequences. 
First, interference between direct and exchange tunneling through the barrier plays already a role 
for nonmagnetic electrodes in transport to lowest order in the tunnel coupling. 
This contrasts with the simpler case of a single tunnel barrier with a magnetic impurity, for which this interference only contributes for ferromagnetic electrodes or in higher order transport.~\cite{kim_electronic_2004,galperin_low-frequency_2004,delgado_spin-transfer_2010,loth_controlling_2010}
Second, we demonstrate that the Coulomb charging energy of the quantum dot can help to perform tunnel spectroscopy on the embedded spin. Even if the excitation energy between the ground state and the first excited state of the impurity spin is larger than any other spin excitation energy, all spin excitation energies are accessible when additional charge states of the dot contribute to transport through the system, which is not possible for the single-barrier case.
Third, we discuss a current-induced switching of the impurity spin. The impurity state influence the spin accumulation on the quantum dot, which in turn acts back on the current through the system, leading to current oscillations as a function of the applied bias. Interestingly, these phenomena occur even for small polarizations of the leads. While the current is only sensitive to the average value of the spin, we find that the zero-frequency Fano factor also contains information about the spin dynamics.
Finally, fourth, we point out how monitoring the exchange field peak in the frequency-dependent Fano factor can detect the switching of the impurity spin for noncollinear magnetizations.

In the second scenario, we consider a $S=1/2$ impurity side coupled to the spin of the electron on the quantum dot. Here, our main aim is to describe the coherent dynamics of the two spins on the quantum dot. We consider the case that the impurity spin is located on the quantum dot, as this allows us to take the exchange coupling between electron and impurity spin into account exactly.
This model can serve to describe different situations.
First, it can describe transport through a quantum dot that is doped with a magnetic atom.
Transport through a quantum dot doped with a single manganese atom has already been studied theoretically. It was shown how the frequency-dependent shot noise can reveal the spin relaxation times.~\cite{contreras-pulido_shot_2010} Furthermore, the electrical control of the manganese spin state as well as the back action of the spin state on transport have been investigated in the absence of Coulomb interaction in the quantum dot.~\cite{fernandez-rossier_single-electron_2007}

Second, our model can be used to describe the coupling of the electron spin on the dot to a nuclear spin via the hyperfine interaction. In general, such a coupling is disadvantageous as it leads to decoherence of the electron spin and therefore can lift, e.g., the Pauli spin blockade in a double quantum dot.~\cite{ono_nuclear-spin-induced_2004,koppens_driven_2006,inarrea_electronic_2007,nadj-perge_disentangling_2010} However, it can also be used to dynamically polarize the nuclear spins in the quantum dot which in turn may be used to control and manipulate the electron spin.~\cite{coish_exchange-controlled_2007,rudner_self-polarization_2007,rudner_electrically_2007,petta_dynamic_2008,inarrea_tunable_2009}

Transport through a quantum dot with a side-coupled spin $1/2$ was discussed in Ref.~\onlinecite{kiesslich_single_2009} for the case of nonmagnetic electrodes. It was shown how to extract the system parameters such as the exchange couplings, the $g$ factors and spin relaxation times from measurements of the current and Fano factor.

The case of noncollinearly magnetized ferromagnetic electrodes was recently investigated by Baumgärtel et al.~\cite{baumgaertel_quadrupole_2010} for a large ferromagnetic exchange interaction. It was shown that in addition to a spin dipole, a spin quadrupole moment accumulates on the quantum dot, driven by a quadrupole current.

In this work, we focus on the opposite regime of small exchange interaction between the side-coupled impurity and electron spin. This situation is particularly suited for the description of the weak hyperfine interaction. We discuss how the frequency-dependent Fano factor can be used to experimentally access the strength of the exchange coupling for large and small external magnetic fields. Furthermore, for the case of a weak external magnetic field, we show how the exchange field acting on the electron spin (but not on the impurity spin) gives rise to a highly nontrivial spin dynamics that manifests itself in the frequency-dependent Fano factor.

Our paper is organized as follows. In Sec.~\ref{sec:Model}, we present the models that describe a magnetic impurity with a large, anisotropic spin localized in one tunnel barrier and a small spin localized on the dot, respectively. We introduce the real-time diagrammatic technique~\cite{kaenig_zero-bias_1996,kaenig_resonant_1996,schoeller_transport_1997,kaenig_quantum_1999} that we use to calculate the transport properties in Sec.~\ref{sec:Technique}. We discuss the form of the reduced density matrix of the quantum dot system and the generalized master equation it obeys for the two systems under investigation in Sec.~\ref{sec:densitymatrix}. Our results for the transport properties are presented in Sec.~\ref{sec:Barrier} for an impurity in the barrier and in Sec.~\ref{sec:Dot} for an impurity on the dot. Finally, we conclude by giving a summary and comparing our results for the two models in Sec.~\ref{sec:Conclusion}.

\section{\label{sec:Model}Model}
In this paper, we consider transport through a quantum-dot spin valve, i.e., a single-level quantum dot tunnel coupled to noncollinearly magnetized ferromagnetic electrodes. We consider additional magnetic impurities either with a large, anisotropic spin localized in the tunnel barrier or with a spin $S=1/2$ on the quantum dot itself. In the following, we define the Hamiltonians for these two cases, respectively.

\subsection{\label{sec:ModelBarrier}Model A: Large spin in the barrier}
\begin{figure}
	\includegraphics[width=.45\textwidth]{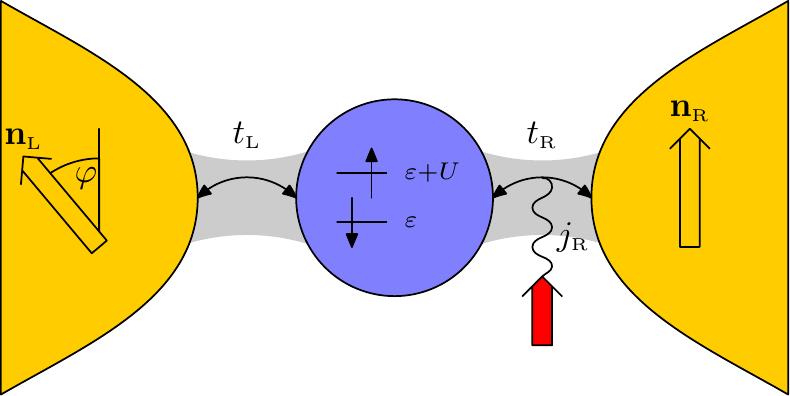}
	\caption{\label{fig:ModelBarrier}(Color online) Schematic of model A, a quantum-dot spin valve with a magnetic impurity localized in the right tunnel barrier.}
\end{figure}
Our model A, which is schematically shown in Fig.~\ref{fig:ModelBarrier}, consists of a quantum-dot spin valve with an impurity embedded in the right tunnel barrier. As already mentioned in Sec.\ref{sec:introduction}, we choose the impurity to be in the tunnel barrier, as this will lead to a simpler conductance pattern, see Sec.~\ref{sec:Barrier} below, which allows to study the spin excitations more easily.
In this case, the Hamiltonian can be written as the sum of four terms describing the two electrodes, the quantum dot, the spin and the tunneling between dot and lead,
\begin{equation}\label{eq:HamiltonianBarrier}
	H=\sum_{r=\text{L,R}}H_r+H_\text{dot}+H_\text{spin}+\sum_{r=\text{L,R}}H_{\text{tun},r}.
\end{equation}
The first term,
\begin{equation}\label{eq:HamiltonianLeads}
	H_r=\sum_{\vec k}\sum_{\sigma=\pm}\varepsilon_{r\vec k\sigma}a_{r\vec k\sigma}^\dagger a_{r\vec k\sigma},
\end{equation}
describes the ferromagnetic electrodes in terms of noninteracting electrons at chemical potential $\mu_r$. We quantize the electron spin in the direction of the magnetization of the respective lead.
The spin polarization is defined as $p_r=(\rho_{r+}-\rho_{r-})/(\rho_{r+}+\rho_{r-})$, where $\rho_{r\sigma}$ is the constant density of states for majority ($\sigma=+$) and minority ($\sigma=-$) spin electrons.

The quantum dot is described by
\begin{equation}
	H_\text{dot}=\sum_{\sigma=\up,\down}\varepsilon c_\sigma^\dagger c_\sigma+Uc_\up^\dagger c_\up c_\down^\dagger c_\down.
\end{equation}
Here, the first term characterizes the single, spin-degenerate quantum dot level with energy $\varepsilon$ measured with respect to the Fermi energy of the leads in equilibrium. The charging energy $U$ is needed to occupy the quantum dot with two electrons. As indicated in Fig.~\ref{fig:ModelBarrier}, we choose the quantization axis of the dot parallel to the magnetization of the right electrode as this simplifies the expressions for the tunnel Hamiltonian, see below.

The third term in Eq.~\eqref{eq:HamiltonianBarrier} describes the magnetic impurity embedded in the right tunnel barrier as a localized spin with Hamiltonian
\begin{equation}\label{eq:HamiltonianSpin}
	H_\text{spin}=-DS_z^2-BS_z.
\end{equation}
We model the spin of magnitude $S$ as having a uniaxial anisotropy $D$ which favors the spin to be in the eigenstates pointing along the $z$ axis. We, furthermore, assume the presence of a magnetic field $B$ acting on the impurity spin. For concreteness, we assume this field to be pointing along the $z$ direction. This choice is motivated by the presence of stray fields from the ferromagnetic electrode which have the tendency to align the impurity along the magnetization of the electrode. As for the quantum dot, we quantize the impurity spin along the direction of the magnetization of the right electrode.

The last part of the Hamiltonian~\eqref{eq:HamiltonianBarrier} describes the tunneling between the dot and the electrodes. For the coupling to the left lead, it takes the form
\begin{multline}\label{eq:HtunL}
	H_\text{tun,L}=\sum_{\vec k} t_\text{L}\left[a_{\text{L}\vec k+}^\dagger\left(\cos\frac{\varphi}{2} c_\up-\sin\frac{\varphi}{2} c_\down\right)\right.\\\left.
	+a_{\text{L}\vec k-}^\dagger\left(\sin\frac{\varphi}{2} c_\up+\cos\frac{\varphi}{2} c_\down\right)\right]+\text{h.c.},
\end{multline}
i.e., majority and minority spin electrons of the leads couple to spin up and spin down states of the quantum dot due to our choice of quantization axes. The coupling to the right lead consists of two terms,
\begin{equation}\label{eq:HtunR}
	H_\text{tun,R}=\sum_{\vec k\sigma} t_\text{R} a_{\text{R}\vec k\sigma}^\dagger c_\sigma + \sum_{\vec k\alpha\beta}j_\text{R} a_{\text{R}\vec k\alpha}^\dagger \vec S\cdot \boldsymbol\sigma_{\alpha\beta}c_\beta+\text{h.c.}.
\end{equation}
Here, $\boldsymbol\sigma_{\alpha\beta}$ denotes the vector of Pauli matrices. The first part describes direct tunneling between the dot and the leads. The second term describes exchange scattering from the impurity spin.

The tunnel matrix elements $t_\text{L}$ and $t_\text{R}$ (which can be chosen real) are related to the spin-dependent tunneling rates via $\Gamma_{r\sigma}=2\pi|t_r|^2\rho_\sigma$. The total tunnel coupling is then given by $\Gamma_r=\sum_{\sigma}\Gamma_{r\sigma}/2$. Similarly, for the exchange tunneling, we relate the corresponding tunneling rate to its (real) matrix element by $J_\text{R}=2\pi|j_\text{R}|^2\frac{\rho_++\rho_-}{2}$. Furthermore, there will be a contribution due to the interference between direct and exchange tunneling through the right barrier. It is characterized by $\eta\sqrt{\Gamma_\text{R}J_\text{R}}$. Here, $\eta=\pm1$ determines the sign of the interference contribution which is governed by the relative sign of $t_\text{R}$ and $j_\text{R}$. The upper sign, $\eta=+1$ applies for equal signs of $t_\text{R}$ and $j_\text{R}$ while the lower sign, $\eta=-1$ applies for different signs of $t_\text{R}$ and $j_\text{R}$.

\subsection{\label{sec:ModelDot}Model B: Small spin on the dot}
\begin{figure}
	\includegraphics[width=.45\textwidth]{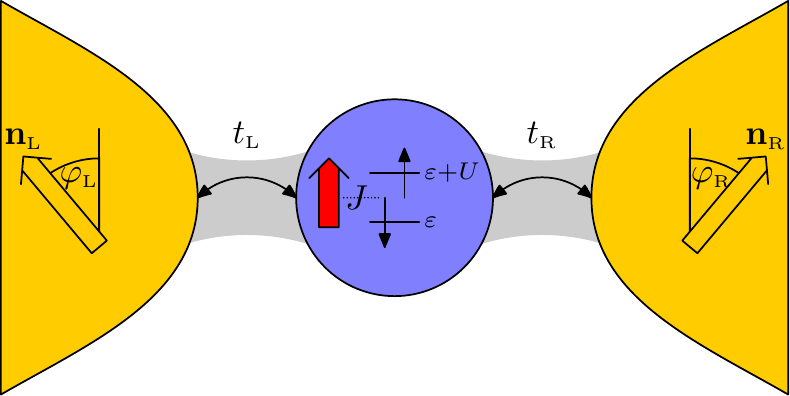}
	\caption{\label{fig:ModelDot}(Color online) Schematic of model B which consists of a quantum-dot spin valve with a magnetic impurity on the quantum dot.}
\end{figure}
Model B consists of a quantum-dot spin valve with an additional spin localized on the quantum dot as shown schematically in Fig.~\ref{fig:ModelDot}.
Here, we restrict ourselves to the case of a $S=1/2$ impurity spin for two reasons. First of all, this keeps the size of the Hilbert space small while still giving rise to the nontrivial spin dynamics we are interested in. Second, for a larger spin that additionally has some anisotropy, spin states would not be degenerate any longer, thereby destroying the possibility to observe the coherent spin dynamics, see also the discussion in Sec.~\ref{sec:Technique}.
The total Hamiltonian now takes the form
\begin{equation}\label{eq:HamiltonianDot}
	H=\sum_{r=\text{L,R}}H_r+H_\text{dot}+H_\text{tun},
\end{equation}
describing the electrodes, the dot containing the impurity spin and the tunneling between dot and leads. The first part, $H_r$, is identical to the one given in Eq.~\eqref{eq:HamiltonianLeads}. For the second term, we have
\begin{multline}\label{eq:Hdot}
	H_\text{dot}=\sum_{\sigma=\up,\down}\varepsilon c_\sigma^\dagger c_\sigma+\frac{B}{2}\left(n_\up-n_\down\right)+U n_\up n_\down\\
	+BS_z+J\sum_{\sigma\sigma'}c_\sigma^\dagger\frac{\vec S\cdot\boldsymbol\sigma_{\sigma\sigma'}}{2}c_{\sigma'},
\end{multline}
The first two terms describe the single dot level with energy $\varepsilon$ and Zeeman splitting $B$ due to an external magnetic field. For simplicity, we assume the magnetic field to point along the quantization axis of the dot which we choose perpendicular to the magnetizations of the leads. The third term describes the Coulomb interaction $U$ which is needed to doubly occupy the quantum dot. The fourth term describes the Zeeman energy of the additional spin on the quantum dot. Here, we assume the same $g$ factor for the electrons on the dot and the impurity spin, for a discussion of a system where electron and impurity spin have different $g$ factors, cf. Ref.~\onlinecite{kiesslich_single_2009}. Finally, the last term describes an exchange interaction between the spin of the electron on the dot and the impurity spin.

The eigenstates of the dot Hamiltonian~\eqref{eq:Hdot} and their corresponding energies are summarized in Table~\ref{tab:dotstates}. The eight states can be classified according to the number of electrons on the dot. For the empty and doubly occupied dot, we have two states each that differ in energy by the Zeeman energy $B$ with the impurity spin pointing up or down. For the singly occupied dot, we have in total four states, three triplet ($S=1$) and one singlet state ($S=0$). While the triplet states are energetically split by the Zeeman energy, singlet and triplet are split by the exchange coupling $J$.
\begin{table}
	\begin{tabular}{|l|l|}
		\hline
		Eigenstate & Energy \\
		\hline
		$\ket{0\up}$ & $E_{0\up}=B/2$ \\
		$\ket{0\down}$ & $E_{0\down}=-B/2$ \\
		\hline
		$\ket{T^+}=\ket{\up\up}$ & $E_{T^+}=\varepsilon+J/4+B$ \\
		$\ket{T^0}=(\ket{\up\down}+\ket{\down\up})/\sqrt{2}$ & $E_{T^0}=\varepsilon+J/4$ \\
		$\ket{T^-}=\ket{\down\down}$ & $E_{T^-}=\varepsilon+J/4-B$ \\
		$\ket{S}=(\ket{\up\down}-\ket{\down\up})/\sqrt{2}$ & $E_{S}=\varepsilon-3/4J$ \\
		\hline
		$\ket{d\up}$ & $E_{d\up}=2\varepsilon+U+B/2$ \\
		$\ket{d\down}$ & $E_{d\down}=2\varepsilon+U-B/2$ \\
		\hline
	\end{tabular}
	\caption{\label{tab:dotstates}Eigenstates of the dot Hamiltonian \eqref{eq:Hdot} and corresponding eigenenergies. The first entry in each ket denotes the state of the quantum dot while the second entry characterizes the impurity spin state.}
\end{table}

The coupling between dot and leads is described by the tunneling Hamiltonian
\begin{multline}\label{eq:Htun}
	H_\text{tun}=\sum_{r\vec k}t_r\left[a_{r\vec k+}^\dagger\left(e^{i\phi_r/2}c_\up +e^{-i\phi_r/2}c_\down\right)\right.\\\left.
	+a_{r\vec k-}^\dagger\left(-e^{i\phi_r/2}c_\up+ e^{-i\phi_r/2}c_\down\right)\right]+\text{h.c.},
\end{multline}
where $\phi_\text{L}=-\phi_\text{R}=\phi/2$ denotes half the angle enclosed by the magnetizations.\cite{braun_theory_2004} We relate the tunnel matrix elements $t_r$ to the tunnel couplings $\Gamma_{r\sigma}$ as for the model discussed in Sec.~\ref{sec:ModelBarrier}. Instead of using the tunnel coupling strength to the left and right lead, we can characterize the dot-lead coupling alternatively by the total tunnel coupling $\Gamma=\Gamma_\text{L}+\Gamma_\text{R}$ and the asymmetry $a=(\Gamma_\text{L}-\Gamma_\text{R})/\Gamma$ with $-1<a<1$.

\section{\label{sec:Technique}Technique}
In order to evaluate the transport properties of the two systems under investigation, we make use of a real-time diagrammatic technique~\cite{kaenig_zero-bias_1996,kaenig_resonant_1996,schoeller_transport_1997,kaenig_quantum_1999} and its extension to systems with noncollinearly magnetized ferromagnetic electrodes.~\cite{kaenig_interaction-driven_2003,braun_theory_2004} The basic idea of this approach is to integrate out the noninteracting degrees of freedom of the electrodes. We, then, arrive at an effective description of the remaining, strongly interacting subsystem in terms of its reduced density matrix $\rho^\text{red}$.

We denote by $P_{\chi_2}^{\chi_1}=\bra{\chi_1}\rho^\text{red}\ket{\chi_2}$ the elements of the reduced density matrix, where $\ket{\chi_1}$ and $\ket{\chi_2}$ are eigenstates of the reduced system. The time evolution of the reduced density matrix elements is governed by a set of generalized master equations,
\begin{equation}\label{eq:MasterEquation}
	\dot P_{\chi_2}^{\chi_1}=-i(\varepsilon_{\chi_1}-\varepsilon_{\chi_2})P_{\chi_2}^{\chi_1}+\sum_{\chi_1'\chi_2'} W_{\chi_2\chi_2'}^{\chi_1\chi_1'}P_{\chi_2'}^{\chi_1'}.
\end{equation}
The first term on the right-hand side describes the coherent evolution of the reduced system. The second term describes the dissipative coupling to the electrodes. The generalized transition rates $W_{\chi_2\chi_2'}^{\chi_1\chi_1'}$ are defined as irreducible self-energies of the dot propagator on the Keldysh contour. They can be evaluated diagrammatically in a perturbation expansion in the tunnel coupling strength $\Gamma$. The corresponding diagrammatic rules are summarized in Appendix~\ref{sec:Rules}.

Expanding the density matrix elements as well as the generalized transition rates in a power series in the tunnel coupling $\Gamma$, we find that in the stationary limit the master equation for the off-diagonal matrix elements to leading order in the tunnel couplings takes the form
\begin{equation}
	0=-i(\varepsilon_{\chi_1}-\varepsilon_{\chi_2})P_{\chi_2}^{\chi_1}
\end{equation}
if $\varepsilon_{\chi_1}-\varepsilon_{\chi_2}\gg\Gamma$. As a consequence, coherent superpositions between states whose energy difference is large compared to the tunnel coupling have to be neglected in the sequential tunneling regime. On the other hand, for superpositions that satisfy $\varepsilon_{\chi_1}-\varepsilon_{\chi_2}\lesssim\Gamma$, the master equation in the stationary limit takes the form
\begin{equation}
	0=-i(\varepsilon_{\chi_1}-\varepsilon_{\chi_2})P_{\chi_2}^{\chi_1}+\sum_{\chi_1'\chi_2'} \left.W_{\chi_2\chi_2'}^{\chi_1\chi_1'}\right|_{\varepsilon_{\chi_1}=\varepsilon_{\chi_2}}P_{\chi_2'}^{\chi_1'}
\end{equation}
to lowest order in the tunnel coupling. Here, the generalized transition rates have to be evaluated at $\varepsilon_{\chi_1}-\varepsilon_{\chi_2}=0$ in order to consistently neglect all effects of order $\Gamma^2$. Hence, in this case, the coherences will not vanish in general.

We define the current through the system as the average of the currents through the left and right tunnel barrier, $I=(I_\text{L}-I_\text{R})/2$. It is given by
\begin{equation}\label{eq:current}
	I=\frac{e}{2\hbar}\vec e^T \vec W^I \vec P.
\end{equation}
Here, we introduced the vector $\vec P$ which contains all density matrix elements written as a vector to allow for a compact notation. The vector $\vec e^T$ is a vector containing a $1$ if the corresponding entry in $\vec P$ is a diagonal density matrix element and a $0$ otherwise. Finally, the quantity $\vec W_r^I$ contains the current rates $W_{\phantom{I}\chi_2\chi_2'}^{I\chi_1\chi_1'}$ which can be obtained diagrammatically similarly to the generalized transition rates $W_{\chi_2\chi_2'}^{\chi_1\chi_1'}$ by replacing one tunneling vertex by a current vertex. The corresponding diagrammatic rules are given in Appendix~\ref{sec:Rules}.

The frequency-dependent current noise is defined as the Fourier transform of the symmetrized current-current correlation function $S=\langle I(t)I(0)+I(0)I(t)\rangle-2\langle I\rangle^2$. In the sequential tunneling regime and for low frequencies, $\omega\lesssim\Gamma$, we can write it as
\begin{multline}\label{eq:finfreqnoise}
	S(\omega)=\frac{e^2}{2\hbar}\vec e^T\left[\vec W^{II}+\vec W^I\left(\boldsymbol\Pi_0^{-1}(\omega)-\vec W\right)^{-1}\vec W^I\right]\vec P\\-2\pi\delta(\omega)\langle I\rangle^2+(\omega\to-\omega),
\end{multline}
where $\vec W^{II}$ is obtained from $\vec W$ by replacing two tunnel vertices by current vertices. The frequency-dependent free dot propagator on the Keldysh contour is given by
\begin{equation}
	\boldsymbol\Pi_0(\omega)^{\chi_1\chi'_1}_{\chi_2\chi'_2}=\frac{i\delta_{\chi_1\chi'_1}\delta_{\chi_2\chi'_2}}{\varepsilon_{\chi_2}-\varepsilon_{\chi_1}-\omega+i0^+}.
\end{equation}
We stress that the frequency-dependent current noise for $\omega\lesssim\Gamma$ is only sensitive to coherences between states with $\varepsilon_{\chi_1}-\varepsilon_{\chi_2}\lesssim\Gamma$, i.e., we can savely neglect all other coherences in the calculation of $S(\omega)$ as in the evaluation of the master equation and the stationary current.
While our discussion of the real-time diagrammatic technique so far was rather general, in the next section, we turn to the explicit form of the density matrix as well as the generalized master equations for the two systems under investigation.

\section{\label{sec:densitymatrix}Reduced density matrix and master equation}
\subsection{\label{ssec:TechniqueBarrier}Model A: Large spin in the barrier}
For a quantum-dot spin valve with a large, anisotropic impurity spin embedded in the tunnel barrier, the eigenstates of the reduced system consisting of quantum dot and impurity spin are products of dot eigenstates $\ket{\chi}\in\{\ket{0},\ket{\up},\ket{\down},\ket{d}\}$ and impurity spin eigenstates $\ket{S_z}\in\{\ket{+S},...,\ket{-S}\}$, $\ket{\xi}=\ket{\chi}\otimes\ket{S_z}$. Assuming the energies of states with different impurity states to differ more than the tunnel coupling, $E_{S_z}-E_{S_z'}\gg\Gamma$, we have to neglect coherent superpositions between states with different impurity states.

The reduced density matrix therefore takes a block diagonal form given by
\begin{equation}
	\rho^\text{red}=\bigoplus_{m=-S}^S \left(
	\begin{array}{cccc}
		P_{0,m} & 0 & 0 & 0 \\
		0 & P_{\up,m} & P_{\down,m}^\up & 0 \\
		0 & P_{\up,m}^\down & P_{\down,m} & 0 \\
		0 & 0 & 0 & P_{d,m}
	\end{array}
	\right).
\end{equation}
In order to give a physically intuitive interpretation of the generalized master equations, we introduce the probabilities to find the dot empty, $P_{0,m}$, singly occupied, $P_{1,m}=P_{\up,m}+P_{\down,m}$, and doubly occupied, $P_{d,m}$, with the impurity in state $\ket{m}$. We collect these quantities in the vector $\vec P_m=(P_{0,m},P_{1,m},P_{d,m})^T$. We furthermore introduce the average spin on the quantum dot $s_{x,m}=\frac{P_{\down,m}^{\up}+P_{\up,m}^{\down}}{2}$, $s_{y,m}=i\frac{P_{\down,m}^{\up}-P_{\up,m}^{\down}}{2}$, and $s_{z,m}=\frac{P_{\up,m}-P_{\down,m}}{2}$. The set of master equations can then be split into one determining the occupation probabilities and one set governing the average dot spin.
In the following, we will keep the time derivative on the left-hand side of the master equations explicitly to allow for a physically intuitive interpretation of the master equations. For the numerical discussion below, these derivatives are equal to zero, however.
The master equations for the occupations are given by
\begin{multline}\label{eq:masterprob}
	\vec{\dot P_m}=
	\vec W_\text{L}^{(0)}\vec P_m+V_\text{L}^{(0)}\vec s_m\cdot \vec n_\text{L}\\
	+\vec W_\text{R}^{(0)}\vec P_m+\vec W_\text{R}^{(+1)}\vec P_{m+1}+\vec W_\text{R}^{(-1)}\vec P_{m-1}\\
	+V_\text{R}^{(0)}\vec s_m\cdot \vec n_\text{R}+V_\text{R}^{(+1)}\vec s_{m+1}\cdot \vec n_\text{R}+V_\text{R}^{(-1)}\vec s_{m-1}\cdot \vec n_\text{R}.
\end{multline}
Here, $\vec W_r^{(\alpha)}$ is a matrix which describes processes that transfer an electron between the dot and lead $r$ and change the state of the impurity spin from $\ket{m}$ to $\ket{m+\alpha}$. Since the impurity is located in the right tunnel barrier, tunneling through the left lead cannot change the impurity state and therefore $\vec W_\text{L}^{(\pm1)}=0$. Changing the impurity state is possible, however, for tunneling through the right barrier which provides a coupling between $\vec P_m$ and $\vec P_{m\pm1}$. Similarly, $V_r^{(\alpha)}$ is a vector which describes the coupling of the occupation probabilities to the spin on the dot, a feature characteristic of a quantum-dot spin valve. Again, we have $V_\text{L}^{(\pm1)}=0$. The precise form of $\vec W_r^{(\alpha)}$ and $V_r^{(\alpha)}$ is given in Appendix~\ref{sec:MasterEquationBarrier}.

The time evolution of the dot spin obeys a Bloch-type equation,
\begin{multline}\label{eq:masterspin}
	\vec{\dot s_m}=\left(\frac{d\vec s_m}{dt}\right)_\text{acc,L}^{(0)}+\left(\frac{d\vec s_m}{dt}\right)_\text{acc,R}^{(0)}+\left(\frac{d\vec s_m}{dt}\right)_\text{acc,R}^{(+1)}\\
	+\left(\frac{d\vec s_m}{dt}\right)_\text{acc,R}^{(-1)}+\left(\frac{d\vec s_m}
	{dt}\right)_\text{rel,L}^{(0)}+\left(\frac{d\vec s_m}{dt}\right)_\text{rel,R}^{(0)}\\
	+\vec s_m\times (\vec B_{m,\text{L}}+\vec B_{m,\text{R}}).
\end{multline}
The first four terms on the right-hand side describe the non-equilibrium spin accumulation on the quantum dot due to the tunneling from and to the spin-polarized leads. Similarly to the master equation for the occupations, we get accumulation terms that change the state of the impurity when tunneling takes place through the right barrier.

The next two terms account for the relaxation of the spin on the dot due to the tunneling out of an electron or the tunneling in of an electron forming a spin singlet with the electron already present on the dot. As these terms arise from generalized transition rates which start and end in a singly occupied state, in the sequential tunneling approximation the state of the impurity spin cannot be changed in these processes. We give the explicit forms of the accumulation and relaxation terms in Appendix~\ref{sec:MasterEquationBarrier}.

The last term describes the precession of the dot spin in an exchange field due to virtual tunneling to the leads. For the coupling to the left lead, we find the usual exchange field~\cite{braun_theory_2004} which is independent of the state of the impurity spin. It is given by
\begin{equation}
	\vec B_{m,\text{L}}=-\vec n_\text{L}\frac{p_\text{L}\Gamma_\text{L}}{\pi}\left(\Phi_\text{L}(\varepsilon)-\Phi_\text{L}(\varepsilon+U)\right),
\end{equation}
where $\Phi_r(x)=\RPG{(x-\mu_r)}$, and $\Psi$ denotes the digamma function. While the first term arises from the spin-dependent level renormalization of an electron virtually tunneling to the lead and back, the second term stems from processes where an electron first tunnels onto the dot and then back into the lead.

\begin{figure}
	\includegraphics[width=\columnwidth]{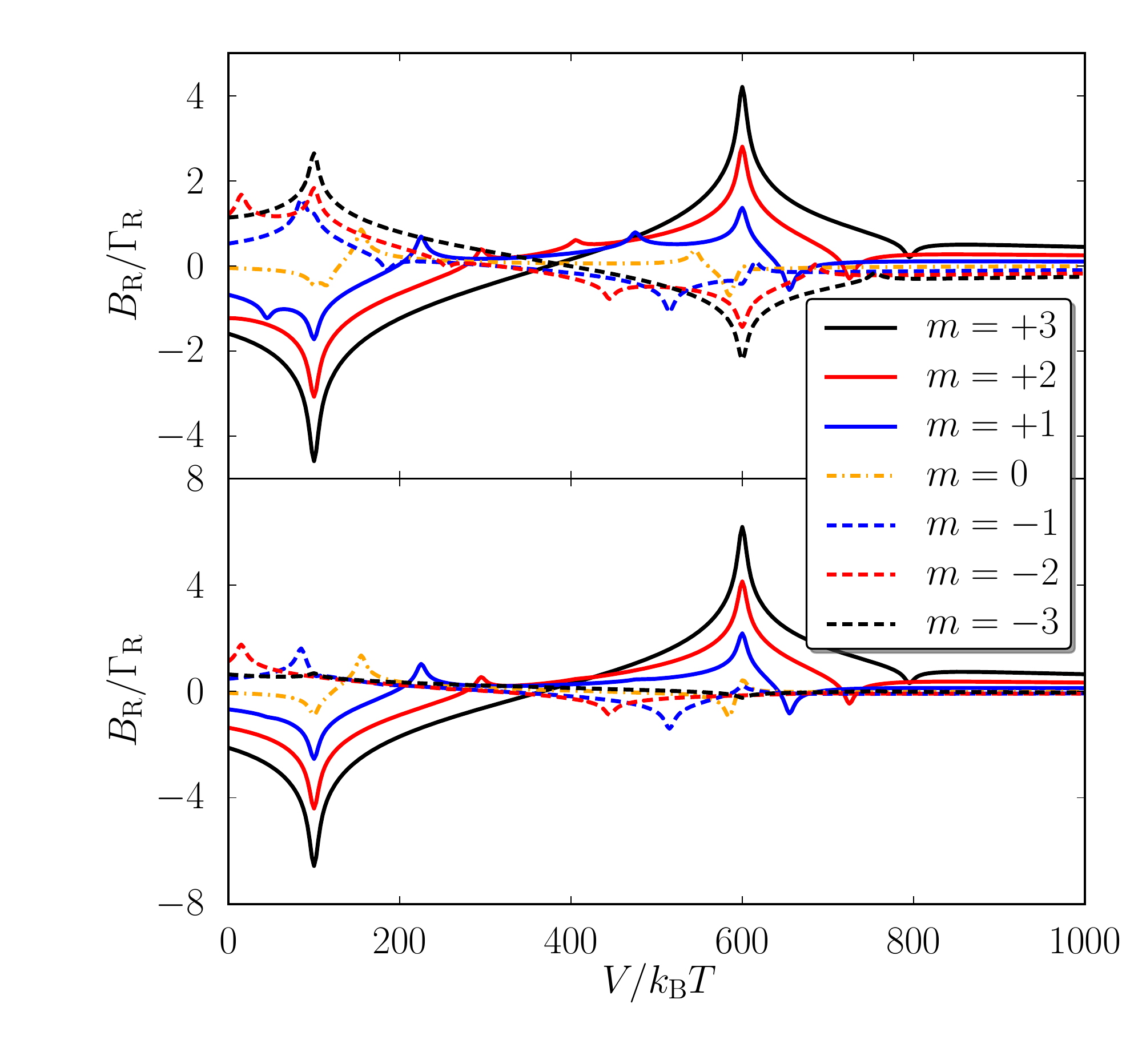}
	\caption{\label{fig:Bexchange}Exchange field due to virtual tunneling through the right barrier as a function of bias voltage $V$ for $p=0.3$ (upper panel) and $p=0.9$ (lower panel). Other parameters are $\varepsilon=-300\kB T$, $U=250\kB T$, $D=17.5\kB T$, $B=10\kB T$, $\Gamma_\text{R}=10J_\text{R}$, $\eta=+1$, $W=500\kB T$, $S=3$.}
\end{figure}
The exchange field due to the coupling to the right lead which points in the direction of $\vec n_\text{R}$ is given by
\begin{widetext}
\begin{multline}
	\vec B_{m,\text{R}}=-\vec n_\text{R}\frac{\tilde\gamma}{\pi}\left(\vphantom{\frac{1}{2}}\Phi_\text{R}(\varepsilon)-\Phi_\text{R}(\varepsilon+U)\right)\\
	+\vec n_\text{R}\frac{A_+(m-1)J_\text{R}}{\pi}\left[\frac{1+p_\text{R}}{2}\Phi_\text{R}(\varepsilon-\alpha_-)+\frac{1-p_\text{R}}{2}\Phi_\text{R}(\varepsilon+U+\alpha_-)-\ln\frac{\beta W}{2\pi}\right]\\
	-\vec n_\text{R}\frac{A_-(m+1)J_\text{R}}{\pi}\left[\frac{1-p_\text{R}}{2}\Phi_\text{R}(\varepsilon+\alpha_+)+\frac{1+p_\text{R}}{2}\Phi_\text{R}(\varepsilon+U-\alpha_+)-\ln\frac{\beta W}{2\pi}\right]
\end{multline}
\end{widetext}
where $\tilde \gamma=p_\text{R}\Gamma_\text{R}+m^2p_\text{R}J_\text{R}+2m\eta\sqrt{\Gamma_\text{R}J_\text{R}}$, $\alpha_\pm=B+(2m\pm1)D$ and $A_\pm(m)=S(S+1)-m(m\pm1)$.

Here, the first term on the right-hand side describes exchange field contributions due to virtual tunneling between dot and lead that does not change the state of the impurity spin. The other two terms are due to virtual tunneling where the intermediate state has an increased/decreased impurity spin state. These processes give rise to a logarithmic divergency of the exchange field cut off by the bandwith $W$ of the lead electrons. To understand this, we consider the impurity in the state $\ket{S_z=S}$. In this case, only the sequences $\ket{\down,S}\to\ket{0,S-1}\to\ket{\down,S}$ and $\ket{\down,S}\to\ket{d,S-1}\to\ket{\down,S}$ of (virtual) transitions are possible while there are no such processes starting from $\ket{\up,S}$. Hence, these processes only renormalize the energy of the spin down state, giving rise to the logarithmic divergency. Similarly, when the impurity is in any other state, the logarithmic contributions to the exchange field do not cancel between processes that increase and decrease the intermediate impurity spin state. As we only consider sequential tunneling, it is clear that our results are only valid if $J_\text{R}\ln\frac{\beta W}{2\pi}\ll\kB T$ and $J_\text{R}\ln\frac{\beta W}{2\pi}\lesssim \Gamma_r$, otherwise higher order logarithmic corrections become important.

We, therefore, find that the presence of the impurity spin in the tunnel barrier has two basic effects on the exchange field. First of all, it alters its strength. Second, due to the presence of the spin-flip processes, it also alters its energy dependence (cf. Fig.~\ref{fig:Bexchange}), giving rise to additional peaks and dips whose separation is governed by the anisotropy $D$, the Zeeman energy $B$, and the size of the impurity spin $S$. Since the transition energies between the various impurity states depend on the states itself, the position of the additional peaks and dips depends on the value of $S_z$.

\subsection{\label{ssec:TechniqueDot}Model B: Small spin on the dot}
\begin{table}
	\begin{tabular}{|l|l|l|}
		\hline
		Case & Parameters & Superpositions of \\
		\hline
		(i) & $B,J,|B-J|\gg\Gamma$ & $-$ \\
		\hline
		(ii) & $B\gg \Gamma$, $J\lesssim\Gamma$ & $\ket{T^0}$, $\ket{S}$ \\
		\hline
		(iii) & $B,J\gg\Gamma$, $|B-J|\lesssim\Gamma$ & $\ket{T^-}$, $\ket{S}$ \\
		\hline
		(iv) & $J\gg\Gamma$, $B\lesssim\Gamma$ & $\ket{0\up}$, $\ket{0\down}$\\
		& & $\ket{T^+}$, $\ket{T^0}$, $\ket{T^-}$\\
		& & $\ket{d\up}$, $\ket{d\down}$\\
		\hline
		(v) & $B,J\lesssim\Gamma$ & $\ket{0\up}$, $\ket{0\down}$\\
		& & $\ket{T^+}$, $\ket{T^0}$, $\ket{T^-}$, $\ket{S}$\\
		& & $\ket{d\up}$, $\ket{d\down}$\\
		\hline
	\end{tabular}
	\caption{\label{tab:superpositions}Coherent superpositions that have to be taken into account in the sequential tunneling regime for different values of the external magnetic field $B$ and the exchange coupling $J$ between the two spins.}
\end{table}
For the case of a $S=1/2$ impurity localized on the quantum dot, the reduced density matrix in the most general case takes the form
\begin{equation}\label{eq:densitymatrix}
	\rho^\text{red}=
	\left(
	\begin{array}{cccccccc}
		P_{0\up} & P_{0\down}^{0\up} & 0 & 0 & 0 & 0 & 0 & 0 \\
		P_{0\up}^{0\down} & P_{0\down} & 0 & 0 & 0 & 0 & 0 & 0 \\
		0 & 0 & P_{T^+} & P_{T^0}^{T^+} & P_{T^-}^{T^+} & P_{S}^{T^+} & 0 & 0 \\
		0 & 0 & P_{T^+}^{T^0} & P_{T^0} & P_{T^-}^{T^0} & P_{S}^{T^0} & 0 & 0 \\
		0 & 0 & P_{T^+}^{T^-} & P_{T^0}^{T^-} & P_{T^-} & P_{S}^{T^-} & 0 & 0 \\
		0 & 0 & P_{T^+}^{S} & P_{T^0}^{S} & P_{T^-}^{S} & P_{S} & 0 & 0 \\
		0 & 0 & 0 & 0 & 0 & 0 & P_{d\up} & P_{d\down}^{d\up} \\
		0 & 0 & 0 & 0 & 0 & 0 & P_{d\up}^{d\down} & P_{d\down} \\
	\end{array}
	\right),
\end{equation}
i.e., apart from the eight diagonal matrix elements that describe the probability to find the system in one of its eigenstates there can be up to 16 coherences. Coherences between states with different electron numbers vanish due to the conservation of total particle number. As discussed above, depending on the energy differences of the states forming the coherent superposition, we either have to take them into account or neglect them in the sequential-tunneling regime. In Table~\ref{tab:superpositions}, we summarize the different transport regimes that arise consequently.

In the following, we will only consider the cases (ii) and (v), i.e., we only consider the case of weak exchange couplings, $J\lesssim\Gamma$. When a large magnetic field is applied, $B\gg\Gamma$, only $S$-$T^0$ coherences have to be taken into account. When the externally applied magnetic field is weak, $B\lesssim\Gamma$, we have to take into account all coherences. There are two reasons focusing on the two cases. On the one hand, they are particularly suited to demonstrate the information about the transport processes contained in the finite-frequency noise. On the other hand, the cases of small exchange couplings are suited to describe the influence of nuclear spins, that couple to the electron spin via hyperfine interaction, on transport through the quantum dot.

\subsubsection{\label{sssec:case2}Case (ii): Large magnetic field}
We first turn to the discussion of the master equation in case (ii) where $B\gg\Gamma$, $J\lesssim\Gamma$. In this case, there are only superpositions of $S$ and $T^0$ present. It is therefore natural to introduce the isospin $\vec I$ via
\begin{align*}
	I_x&=\frac{P^S_{T^0}+P^{T^0}_S}{2},& I_y&=i\frac{P^S_{T^0}-P^{T^0}_S}{2},\\ I_z&=\frac{P_{T^0}-P_S}{2},& I_0&=P_{T^0}+P_S,
\end{align*}
to bring the master equation into a physically intuitive form. Similar to the case of an ordinary quantum-dot spin valve,~\cite{braun_theory_2004} we can now split the master equation into one set governing the occupation probabilities that we summarize in the vector $\vec P=(P_{0\up},P_{0\down},P_{T^+},I_0,P_{T^-},P_{d\up},P_{d\down})$ and one set governing the time evolution of the isospin $\vec I=(I_x,I_y,I_z)$. However, there is an important difference. While in the ordinary quantum-dot spin valve, there is a real spin accumulating on the quantum dot, here we have an isospin accumulation as a real spin accumulation is suppressed by the large external magnetic field. The master equation of the occupation probabilities reads
\begin{equation}\label{eq:MEP}
	\dot{\vec P}=\vec W\cdot \vec P+\vec V(\vec I\cdot\vec e_x),
\end{equation}
where $\vec W$ denotes a matrix that contains the golden rule transition rates between the various dot states and $\vec V$ is a vector that characterizes the influence of the isospin on the dot occupation whose precise form is given in Appendix~\ref{sec:MasterEquationDot}.

The master equation that governs the time evolution of the isospin is given by
\begin{equation}\label{eq:MEI}
	\dot{\vec I}=\left(\frac{d\vec I}{dt}\right)_\text{acc}+\left(\frac{d\vec I}{dt}\right)_\text{relax}+\vec I\times\sum_r\vec B_r.
\end{equation}
Here, the first term,
\begin{widetext}
\begin{multline}\label{eq:SpinAccumulation}
	\left(\frac{d\vec I}{dt}\right)_\text{acc}=\frac{1}{2}\sum_r\Gamma_r\left\{-f_r^+(\varepsilon-B/2)P_{0\up}+f_r^+(\varepsilon+B/2)P_{0\down}-f_r^+(\varepsilon+U+B/2)P_{d\up}+f_r^-(\varepsilon+U-B/2)P_{d\down}\phantom{\frac{1}{2}}\vphantom{\frac{1}{2}}\right.\\\left.
	+\frac{1}{2}\left[f_r^-(\varepsilon-B/2)-f_r^-(\varepsilon+B/2)+f_r^+(\varepsilon+U+B/2)-f_r^+(\varepsilon+U-B/2)\vphantom{\frac{1}{2}}\right]I_0\right\}\vec e_x,
\end{multline}
\end{widetext}
describes the accumulation of the isospin along the $x$ axis due to electrons tunneling onto and off the dot. Similarly, the second term describes a relaxation of the isospin
\begin{multline}\label{eq:SpinRelaxation}
	\left(\frac{d\vec I}{dt}\right)_\text{relax}=-\frac{1}{2}\sum_r\Gamma_r\left[f_r^-(\varepsilon-B/2)+f_r^-(\varepsilon+B/2)\vphantom{\frac{1}{2}}\right.\\\left.
	+f_r^+(\varepsilon+U-B/2)+f_r^+(\varepsilon+U+B/2)\vphantom{\frac{1}{2}}\right]\vec I.
\end{multline}
Finally, the last term describes the precession of the isospin in an exchange field that is given by
\begin{align}\label{eq:ExchangeField}
	B_{rx}&=\frac{\Gamma_r}{2\pi}\left(\Phi_r(\varepsilon-B/2)-\Phi_r(\varepsilon+B/2)\vphantom{\frac{1}{2}}\right.\notag\\
	&\left.
	\phantom{=}+\Phi_r(\varepsilon+U+B/2)-\Phi_r(\varepsilon+U-B/2)\vphantom{\frac{1}{2}}\right)\notag\\
	B_{ry}&=0\\
	B_{rz}&=J\notag
\end{align}
The exchange field describes the level splitting between $\ket{T^0}$ and $\ket{S}$ which is due to the finite exchange coupling $J$ as well as due to virtual tunneling processes that renormalize the energies of the two states in a different way. As can be inferred from Eq.~\eqref{eq:MEI}, it gives rise to a precession of the accumulated isospin around the exchange field.

The current through the quantum dot is given by
\begin{widetext}
\begin{multline}\label{eq:Current}
	I=\frac{\Gamma_\text{L}}{2}\left[\left(\vphantom{\frac{1}{2}}f^+_\text{L}(\varepsilon+B/2)+f^+_\text{L}(\varepsilon-B/2)\right)\left(\vphantom{\frac{1}{2}}P_{0\up}+P_{0\down}\right)-\left(\vphantom{\frac{1}{2}}f^-_\text{L}(\varepsilon+U+B/2)+f^-_\text{L}(\varepsilon+U-B/2)\right)\left(\vphantom{\frac{1}{2}}P_{d\up}+P_{d\down}\right)\right.\\\left.
	-\left(\vphantom{\frac{1}{2}}f^-_\text{L}(\varepsilon+B/2)-f^+_\text{L}(\varepsilon+U-B/2)\right)P_{T^+}-\left(\vphantom{\frac{1}{2}}f^-_\text{L}(\varepsilon-B/2)-f^+_\text{L}(\varepsilon+U+B/2)\right)P_{T^-}\right.\\\left.
	-\frac{1}{2}\left(\vphantom{\frac{1}{2}}f^-_\text{L}(\varepsilon+B/2)+f^-_\text{L}(\varepsilon-B/2)-f^+_\text{L}(\varepsilon+U+B/2)-f^+_\text{L}(\varepsilon+U-B/2)\right)I_0\right.\\\left.
	-\left(\vphantom{\frac{1}{2}}f^-_\text{L}(\varepsilon+B/2)-f^-_\text{L}(\varepsilon-B/2)+f^+_\text{L}(\varepsilon+U+B/2)-f^+_\text{L}(\varepsilon+U-B/2)\right)I_x\right]-\left(\text{L}\to\text{R}\right).
\end{multline}
\end{widetext}
It depends on the occupation probabilities as well as on the $x$ component of the accumulated isospin. This resembles the normal quantum-dot spin valve where the current also depends on both, the dot occupations as well as on the dot spin.~\cite{braun_theory_2004}

\subsubsection{\label{sssec:case5}Case (v): Small magnetic field}
We now turn to the discussion of the master equation in the case $B,J\lesssim\Gamma$. In this case, we have to include all coherences of the reduced density matrix, Eq.~\eqref{eq:densitymatrix}.

To allow for a physical interpretation of the different matrix elements, we introduce the probabilities to find the quantum dot empty, $P_0$, singly occupied, $P_1$, and doubly occupied, $P_d$. Furthermore, we introduce the expectation values of the electron spin on the dot, $\vec S_1$, as well as the expectation values for the impurity spin when the dot is empty, $\vec S_0$, singly occupied, $\vec S_2$, and doubly occupied, $\vec S_d$. While for a single spin $1/2$ the description of its density matrix in terms of spin expectation values is sufficient, this is in general no longer true for a system of two spin $1/2$ particles.~\cite{baumgaertel_quadrupole_2010} For the case of small magnetic fields that we consider here, we therefore have to introduce in addition the expectation values of the scalar product between electron and impurity spin, $\vec S_1\cdot\vec S_2$, and their vector product, $\vec S_1\times\vec S_2$. Finally, we also need to introduce the quadrupole moment~\cite{edmunds_angular_1996,baumgaertel_quadrupole_2010}
\begin{equation}
	Q_{ij}=\frac{1}{2}\left(S_{1i}S_{2j}+S_{1j}S_{2i}\right)-\frac{1}{3}\vec S_1\cdot\vec S_2\delta_{ij}.
\end{equation}
The quadrupole moment is a symmetric tensor, $Q_{ij}=Q_{ji}$. Its diagonal elements are not independent of each other as they satisfy the sum rule $\sum_i Q_{ii}=0$, i.e., the trace of $\vec Q$ vanishes.

In Appendix~\ref{app:densitymatrix} we give the explicit expressions that relate the above quantities to the density matrix elements in Eq.~\eqref{eq:densitymatrix}. We note that in the case $B\gg\Gamma$, $J\lesssim\Gamma$ where we only have taken into account the $S$-$T^0$ superpositions, we could have expressed the reduced density matrix in terms of the quantities just introduced as well. However, by choosing a description in terms of the isospin, we obtain a much simpler master equation.

Using the physical quantities we just discussed, we can split the master equation into several sets. The first set
\begin{widetext}
\begin{equation}\label{eq:occupations}
	\frac{d}{dt}
	\left(
	\begin{array}{c}
		P_0 \\
		P_1 \\
		P_d \\
	\end{array}
	\right)
	=\sum_r\Gamma_r
	\left(
	\begin{array}{ccc}
		-2 f_r^+(\varepsilon) & f_r^-(\varepsilon) & 0 \\
		2 f_r^+(\varepsilon) & -f_r^-(\varepsilon)-f_r^+(\varepsilon+U) & 2f_r^-(\varepsilon+U) \\
		0 & f_r+(\varepsilon+U) & -2f_r^-(\varepsilon+U)
	\end{array}
	\right)
	\left(
	\begin{array}{c}
		P_0 \\
		P_1 \\
		P_d \\
	\end{array}
	\right)
	+\sum_r 2p_r\Gamma_r
	\left(
	\begin{array}{c}
		f_r^-(\varepsilon) \\
		-f_r^-(\varepsilon)+f_r^+(\varepsilon+U) \\
		-f_r^+(\varepsilon+U)
	\end{array}
	\right)
	\vec S_1\cdot \vec n_r,
\end{equation}
\end{widetext}
describes the evolution of the occupation probabilities. It takes a form identical to the case of a normal quantum-dot spin valve, i.e., it exhibits a coupling of the occupations to the spin accumulated on the quantum dot. Interestingly, the occupations do not couple neither to the impurity spin, the scalar or vector product of $\vec S_1$ and $\vec S_2$ nor to the quadrupole moments directly. They are only influenced by these quantities due to their influence on the accumulated electron spin $\vec S_1$.

The equation governing the time evolution of the electron spin in the dot is given by
\begin{widetext}
\begin{equation}\label{eq:spinelectron}
	\frac{d\vec S_1}{dt}=\sum_r\left[p_r\Gamma_r\left(f_r^+(\varepsilon)P_0-\frac{f_r^-(\varepsilon)-f_r^+(\varepsilon+U)}{2}P_1-f_r^-(\varepsilon+U)P_d\right)\vec n_r-\frac{\vec S_1}{\tau_r}-\vec S_1\times\vec B_{r,\text{ex}}\right]-\vec S_1\times\vec B+J(\vec S_1\times\vec S_2),
\end{equation}
\end{widetext}
where $1/\tau_r=\Gamma_r\left(f_r^-(\varepsilon)+f_r^+(\varepsilon+U)\right)$ and $\vec B_{r,\text{ex}}=-\frac{p\Gamma_r}{\pi}\left(\Phi_r(\varepsilon)-\Phi_r(\varepsilon+U)\right)\vec n_r$ is the usual exchange field acting on the electron spin accumulated on the dot. Again, we find a strong similarity to the case of the normal quantum-dot spin valve. While the first term in brackets describes the accumulation of spin on the dot along $\vec n_r$ due to spin-dependent tunneling of electrons between dot and leads, the second term describes a relaxation of the dot spin due to tunneling. The third term in brackets describes the precession of the dot spin in the exchange field generated by virtual tunneling between dot and leads. The last two terms finally describe the influence of an external magnetic field and the exchange coupling to the impurity spin.

The master equations for the impurity spin in the presence of zero, one and two electrons on the dot can be written as
\begin{widetext}
\begin{multline}\label{eq:spinimpurity}
	\frac{d}{dt}
	\left(
	\begin{array}{c}
		\vec S_0 \\
		\vec S_2 \\
		\vec S_d
	\end{array}
	\right)
	=
	\sum_r \Gamma_r
	\left(
	\begin{array}{ccc}
	-2f_r^+(\varepsilon) & f_r^-(\varepsilon) & 0 \\
	2f_r^+(\varepsilon) & -f_r^-(\varepsilon)-f_r^+(\varepsilon+U) & 2f_r^-(\varepsilon+U) \\
	0 & f_r^+(\varepsilon+U) & -2f_r^-(\varepsilon+U)
	\end{array}
	\right)
	\left(
	\begin{array}{c}
		\vec S_0 \\
		\vec S_2 \\
		\vec S_d
	\end{array}
	\right)\\
	+\sum_r 2p_r\Gamma_r
	\left(
	\begin{array}{c}
		f_r^-(\varepsilon) \\
		-f_r^-(\varepsilon)+f_r^+(\varepsilon+U) \\
		-f_r^+(\varepsilon+U)
	\end{array}
	\right)
	\left[\left(\vec Q+\frac{1}{3}\vec S_1\cdot\vec S_2\right)\cdot \vec n_r+\frac{1}{2}\left(\vec S_1\times \vec S_2\right)\times \vec n_r\right]\\
	-\left(
	\begin{array}{c}
		\vec S_0 \\
		\vec S_2 \\
		\vec S_d
	\end{array}
	\right)
	\times \vec B
	+J\left(
	\begin{array}{c}
		0 \\
		\vec S_1\times\vec S_2 \\
		0
	\end{array}
	\right).
\end{multline}
\end{widetext}
Here, the first term on the right-hand side describes transitions between the three quantities by tunneling of electrons in analogy to the first term in the equation for the occupations, Eq.~\eqref{eq:occupations}. The second term characterizes the coupling to the quadrupole moments as well as the scalar and vector product of $\vec S_1$ and $\vec S_2$. This resembles the coupling of the dot occupations to the electron spin on the dot in Eq.~\eqref{eq:occupations}. Finally, the terms in the third line describe the precession of the impurity spin in an externally applied magnetic field as well as the influence of the exchange interaction between electron and impurity spin.

The master equations governing the time evolution of the scalar and vector product between the electron and impurity spin are given by
\begin{widetext}
\begin{multline}\label{eq:spinvector}
	\frac{d}{dt}(\vec S_1\times\vec S_2)=\sum_r\left[-p_r\Gamma_r\left(f_r^+(\varepsilon)\vec S_0-\frac{f_r^-(\varepsilon)-f_r^+(\varepsilon+U)}{2}\vec S_2-f_r^-(\varepsilon+U)\vec S_d\right)\times\vec n_r-\frac{\vec S_1\times\vec S_2}{\tau_r}\right.\\\left.
	+\left(\vec Q-\frac{2}{3}(\vec S_1\cdot\vec S_2)\right)\cdot\vec B_{r,\text{ex}}\right]
	-(\vec S_1\times\vec S_2)\times\vec B+\frac{J}{2}\left(\vec S_1-\vec S_2\right),
\end{multline}
\begin{equation}\label{eq:spinscalar}
	\frac{d}{dt}(\vec S_1\cdot\vec S_2)=\sum_r\left[p_r\Gamma_r\left(f_r^+(\varepsilon)\vec S_0-\frac{f_r^-(\varepsilon)+f_r^+(\varepsilon+U)}{2}\vec S_2-f_r^-(\varepsilon+U)\vec S_d\right)\cdot\vec n_r-\frac{\vec S_1\cdot\vec S_2}{\tau_r}+(\vec S_1\times\vec S_2)\cdot \vec B_{r,\text{ex}}\right].
\end{equation}
\end{widetext}
Their structure closely resembles Eq.~\eqref{eq:spinelectron} in that there are terms which describe the accumulation, relaxation, and the influence of the spin precession due to the exchange field. Furthermore, the vector product turns out to be sensitive to an external magnetic field as well as to the exchange coupling between the spins.

Finally, the master equation for the quadrupole moment takes the form
\begin{widetext}
\begin{multline}\label{eq:quadrupole}
	\frac{d}{dt}\vec Q_{ij}=\sum_r\left[p_r\Gamma_r f_r^+(\varepsilon)\left(\frac{1}{2}\left(\vec S_{0,i}\vec n_{r,j}+\vec S_{0,j}\vec n_{r,i}\right)-\frac{1}{3}(\vec S_0\cdot\vec n_r)\delta_{ij}\right)\right.\\\left.
	-p_r\Gamma_r\frac{f_r^-(\varepsilon)-f_r^+(\varepsilon+U)}{2}\left(\frac{1}{2}\left(\vec S_{2,i}\vec n_{r,j}+\vec S_{2,j}\vec n_{r,i}\right)-\frac{1}{3}(\vec S_2\cdot\vec n_r)\delta_{ij}\right)\right.\\\left.
	-p_r\Gamma_r f_r^-(\varepsilon+U)\left(\frac{1}{2}\left(\vec S_{d,i}\vec n_{r,j}+\vec S_{d,j}\vec n_{r,i}\right)-\frac{1}{3}(\vec S_d\cdot\vec n_r)\delta_{ij}\right)\right.\\\left.
	-\frac{\mathbf Q_{ij}}{\tau_r}
	-\frac{1}{2}\left(\frac{1}{2}(\vec S_1\times\vec S_2)_i(\vec B_{r,\text{ex}})_j+\frac{1}{2}(\vec S_1\times\vec S_2)_j(\vec B_{r,\text{ex}})_i-\frac{1}{3}(\vec S_1\times\vec S_2)\cdot\vec B_{r,\text{ex}}\delta_{ij}\right)\right.\\\left.
	-\frac{1}{2}\varepsilon_{ilm}\vec Q_{lj} (\vec B_{r,\text{ex}})_m -\frac{1}{2}\varepsilon_{jlm}\vec Q_{li} (\vec B_{r,\text{ex}})_m
	\right]
	-\varepsilon_{ilm}\vec Q_{lj}\vec B_m-\varepsilon_{jlm}\vec Q_{li}\vec B_m.
\end{multline}
\end{widetext}
The first three terms on the right-hand side describe the accumulation of quadrupole moment on the quantum dot. Similarly, the fourth term is related to the relaxation of the quadrupole moment. Finally, the other terms describe the precesional motion of the quadrupole moment in the exchange field as well as due to an external magnetic field.

The current through tunnel barrier $r$ is given by
\begin{widetext}
\begin{equation}
	I_r=\Gamma_rf_r^+(\varepsilon)P_0-\Gamma_r\frac{f_r^-(\varepsilon)-f_r^+(\varepsilon+U)}{2}P_1-\Gamma_rf_r^-(\varepsilon+U)P_d-p\Gamma_r\left[f_r^-(\varepsilon)+f_r^+(\varepsilon+U)\vphantom{\frac{1}{2}}\right]\vec S_1\cdot\vec n_r.
\end{equation}
\end{widetext}
Although this is precisely the same form as for the normal quantum-dot spin valve, the current nevertheless contains information about the nontrivial spin dynamics on the dot, as the master equation for the dot spin couples to the other density matrix elements.

\section{\label{sec:Barrier}Results - Large spin in the barrier}
In this section we discuss the transport properties of a quantum-dot spin valve with a large, anisotropic impurity spin located in the right tunnel barrier. We will focus our attention on systems that are symmetric in the sense that $p_\text{L}=p_\text{R}\equiv p$ and $\Gamma_\text{L}=\Gamma_\text{R}\equiv \Gamma/2$. Furthermore we assume the bias voltage to be applied symmetrically, $V_\text{L}=-V_\text{R}=-V/2$.

\subsection{\label{ssec:currentcoll}Collinear magnetizations}
In the following, we are going to discuss transport for collinear magnetizations. Most of the time we will restrict ourselves to the case of parallel magnetizations as we focus on moderate polarizations where the effects due to the impurity spin are much more important than the effects due to the relative orientation of the leads.

\subsubsection{\label{sssec:interference}Interference between direct and exchange tunneling}
From the form of the tunneling Hamiltonian~\eqref{eq:HtunR} it is obvious that interference can take place between electrons tunneling directly through the barrier and electrons experiencing an exchange interaction with the impurity spin. For transport through a single tunnel barrier containing a localized spin~\cite{appelbaum_s-d_1966,appelbaum_exchange_1967,kim_electronic_2004,misiorny_magnetic_2007,sothmann_nonequilibrium_2010}, the interference contributions cancel between the spin-up and spin-down channel. Only for ferromagnetic leads~\cite{galperin_low-frequency_2004,delgado_spin-transfer_2010,loth_controlling_2010,fransson_theory_2010} or in the presence of spin-orbit interactions, one is sensitive to the interference terms.

This is different for the system under investigation here. We find that the interference terms influence the current even for unpolarized leads.
In contrast to the single-barrier case where we just have to sum up the contributions from spin-up and spin-down electrons to the current, in the quantum dot case we have to separately compare the rates for tuneling in and out of the dot for each spin direction. While in the nonmagnetic case equal amounts of spin-up and spin-down electrons enter the dot from the left lead, the rates for leaving the dot are different due to the interference terms. This in turn gives rise to a spin accumulation on the quantum dot which reduces the current through the quantum dot.

Unfortunately, in our system there is no way to tune the phase of the interference terms experimentally as is possible, e.g., in an Aharonov-Bohm interferometer and thereby check the influence of the interference terms on the current. Nevertheless, it should be possible to detect the presence of the interference term experimentally and to detect its sign. Approximating the Fermi functions as step functions which is reasonable away from the threshold voltages, we can calculate the current through the system analytically in the various transport regions.

We first consider the case of unpolarized leads, $p_r=0$. In this case, the current in region I where transport takes place through the singly and doubly occupied dot (cf. Fig.~\ref{fig:conductance}) is given by
\begin{equation}\label{eq:currentI}
	I_\text{I}=\frac{2\Gamma_\text{L}(\Gamma_\text{R}+S^2J_\text{R})}{\Gamma_\text{L}+2(\Gamma_\text{R}+S^2J_\text{R})},
\end{equation}
i.e. it is sensitive to the couplings and the size of the barrier spin but not to the interference term. Similarly, in region II where spin excitations become possible, the current turns out to be insensitive to the interference term,
\begin{equation}\label{eq:currentII}
	I_\text{II}=\frac{2\Gamma_\text{L}(\Gamma_\text{R}+S(S+1)J_\text{R})}{\Gamma_\text{L}+2(\Gamma_\text{R}+S(S+1)J_\text{R})}.
\end{equation}
This is different in region III where transport takes place through the empty and singly-occupied dot but the spin cannot be excited. Here, the current is given by
\begin{equation}\label{eq:currentIII}
	I_\text{III}=\frac{2\Gamma_\text{L}((\Gamma_\text{R}+S^2J_\text{R})^2-4S^2\sqrt{\Gamma_\text{R}J_\text{R}}^2)}{(\Gamma_\text{R}+S^2J_\text{R})(2\Gamma_\text{L}+\Gamma_\text{R}+S^2J_\text{R})-4S^2\sqrt{\Gamma_\text{R}J_\text{R}}^2)},
\end{equation}
i.e., the current now also depends on the interference term. From Eq.~\eqref{eq:currentIII}, we infer that for $\Gamma_\text{R}=SJ_\text{R}$ the current vanishes exactly in region III. Equation~\eqref{eq:currentIII} also shows that the current in region III is only sensitive to the absolute value of the interference term but not to its  sign.

This is different in the regime where the empty and singly occupied dot contribute to transport and spin excitations are possible. As the analytic result for the current in this regime is rather lengthy, we do not give it here. Instead, we now focus on transport for parallely magnetized leads. In region I, the current is now given by
\begin{equation}\label{eq:currentIP}
	I_\text{I}=\frac{2\Gamma_\text{L}\left(\Gamma_\text{R}+S^2J_\text{R}+2pS\eta\sqrt{\Gamma_\text{R}J_\text{R}}\right)}{\Gamma_\text{L}+2\left(\Gamma_\text{R}+S^2J_\text{R}\right)}.
\end{equation}
Here, the current is clearly sensitive to the sign of the interference term which provides a way to access it in experiments. Similar expressions for the current in regions II and III can be found for parallely magnetized leads. As these expressions are rather lengthy, we do not give them here.

\subsubsection{\label{sssec:spectroscopy}Spin spectroscopy}
\begin{figure}
	\includegraphics[width=\columnwidth]{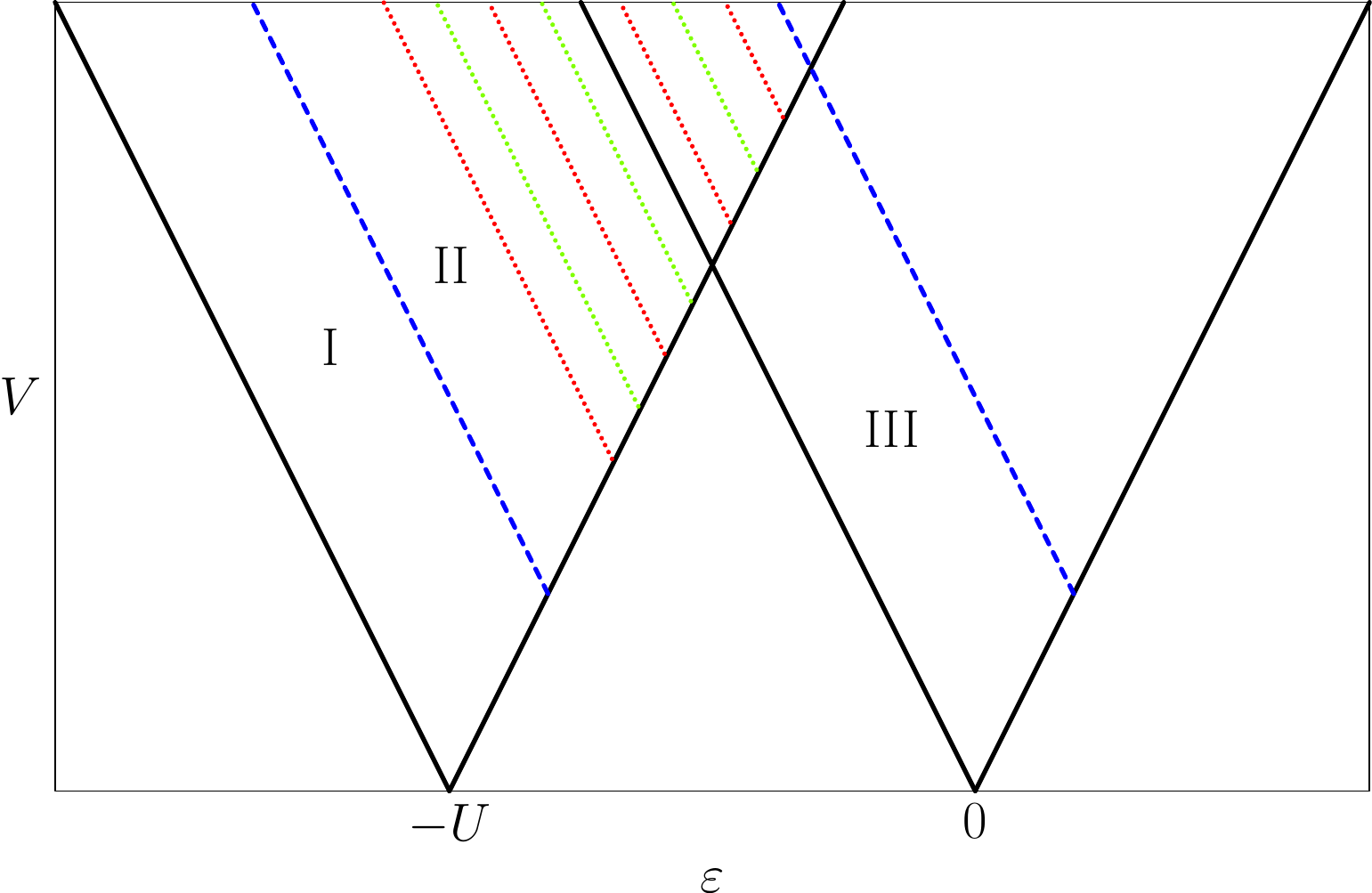}
	\caption{\label{fig:conductance}(Color online) Schematic of the differential conductance as a function of level position and applied bias voltage. Thick black lines mark the onset of transport through the dot. Blue dashed lines indicate the onset of impurity excitations. Dashed lines mark the gaining of energy from the impurity to allow all three charge states of the dot.}
\end{figure}
Inelastic spin tunneling spectroscopy~\cite{heinrich_single-atom_2004,hirjibehedin_spin_2006,hirjibehedin_large_2007,otte_role_2008,otte_spin_2009,parks_mechanical_2010,zyazin_electric_2010,fernandez-rossier_theory_2009,fransson_spin_2009,persson_theory_2009,lorente_efficient_2009,delgado_spin-transfer_2010,sothmann_nonequilibrium_2010} allows to determine the spectrum of a spin embedded in a tunnel barrier by studying the steps in the differential conductance that occur whenever an inelastic transport channel opens up. However, for a simple spin Hamiltonian of the form~\eqref{eq:HamiltonianSpin} this does not allow to determine the parameters $D$ and $B$ separately. In this case, the energy difference between the ground state and the first excited state is larger than all other excitation energies.~\cite{jo_signatures_2006,zyazin_electric_2010} In consequence, as soon as the system can be brought into the first excited state, all other excited states can also be reached energetically. Hence, there would be only a signal at $\Delta=E_{S-1}-E_{S}=(2S-1)D+B$.

This is different for the system additionally containing a quantum dot between the electrodes, as is shown schematically in Fig.~\ref{fig:conductance}.
For level positions $\varepsilon<-U$, the dot is doubly occupied when no bias is applied. Upon increasing the bias, we enter region I where transport takes place through the singly and doubly occupied dot. When increasing the bias voltage above the blue dashed line, exciting the spin becomes possible, similarly as in a single barrier discussed above which only provides information about a linear combination of $D$ and $B$. However, upon increasing the bias further, we reach the red dotted line. At this point, the electron on the dot with an energy below the right Fermi level can gain enough energy to leave the dot to the right lead by changing the impurity state from $S_z=S-1$ to $S_z=S$. As this opens up a new transport channel, the onset of this process yields a signal in the differential conductance. Similarly, at the next yellow line the process $\ket{\down}\otimes\ket{-S+1}\to\ket{0}\otimes\ket{-S}$ becomes possible, again giving rise to a conductance signal. This scheme continues for all transitions of the impurity spin which are characterized by transition energies $(2S_z-1)D+B$ for $S_z>0$ and $(-(2S_z+1)D-B)$ for $S_z<0$. When we finally reach the thick black line, the empty dot state can also be reached by ordinary tunneling events. For even larger bias voltages we find another series of conductance signals which are now associated with electrons in the lower level which become able to excite the impurity when leaving to the right lead. The important difference to the small bias case discussed above is that now transport through the upper level can bring the impurity spin into all excited states such that transitions between these states also are all visible.

For unpolarized leads, the conductance pattern discussed above allows to determine $D$ and the absolute value of $B$. As no spatial direction is distinguished, there is no possibility to determine also the sign of $B$. This is different for polarized leads where the spatial symmetry is broken by the magnetizations. In this case, one finds that the differential conductance shows an alternating pattern of positive and negative differential conductance which depends on the sign of $B$. We discuss the mechanism leading to this behavior in the following.

\subsubsection{\label{sssec:cips}Current-induced switching and spin-dependent transport}
\begin{figure}
	\includegraphics[width=\columnwidth]{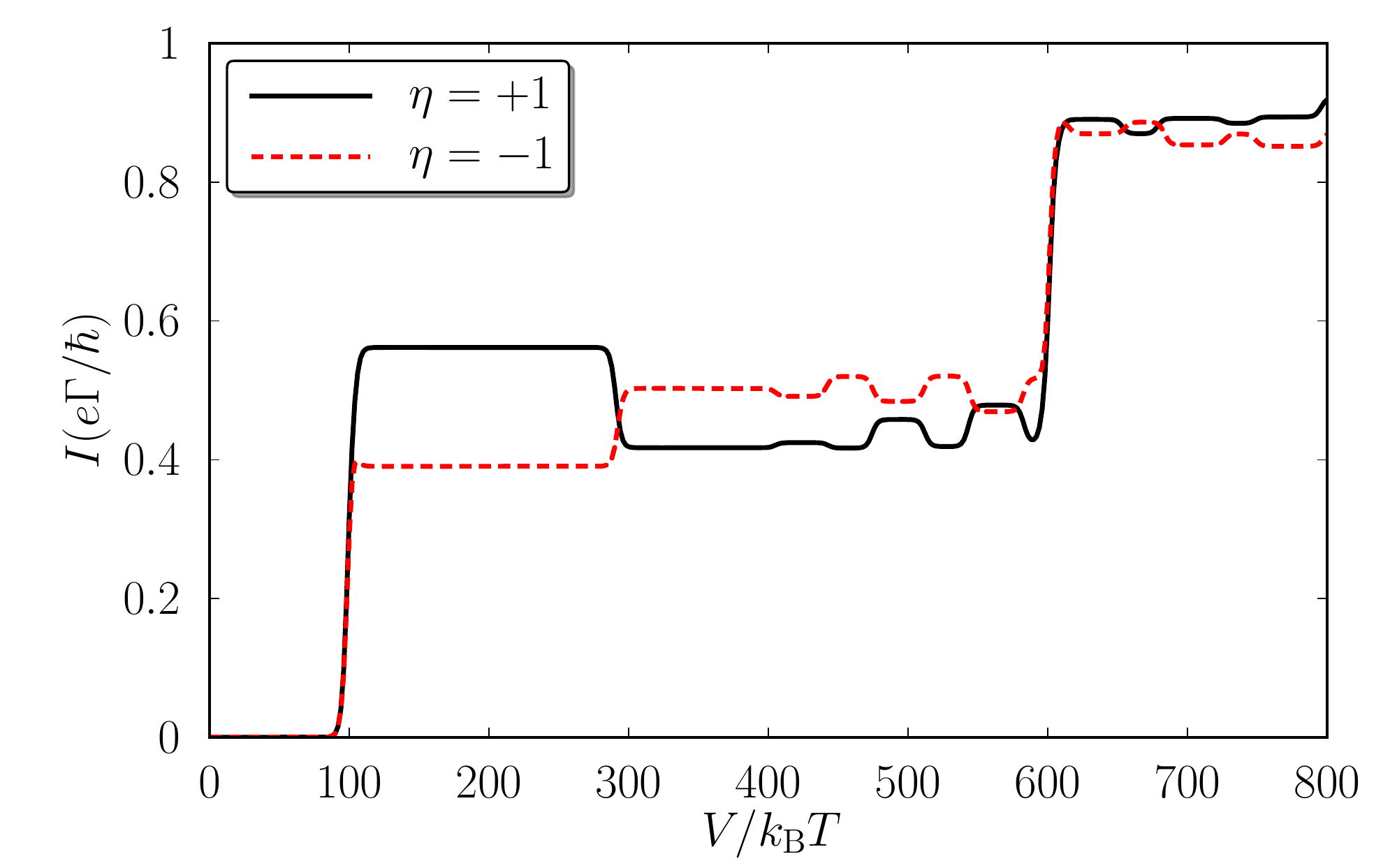}
	\caption{\label{fig:current}(Color online) Current as a function of bias voltage for parallel magnetizations, $p=0.3$ and $\Gamma_\text{L}=\Gamma_\text{R}=J_\text{R}$. Other parameters are the same as in Fig. \ref{fig:Bexchange}.}
\end{figure}
\begin{figure}
	\includegraphics[width=\columnwidth]{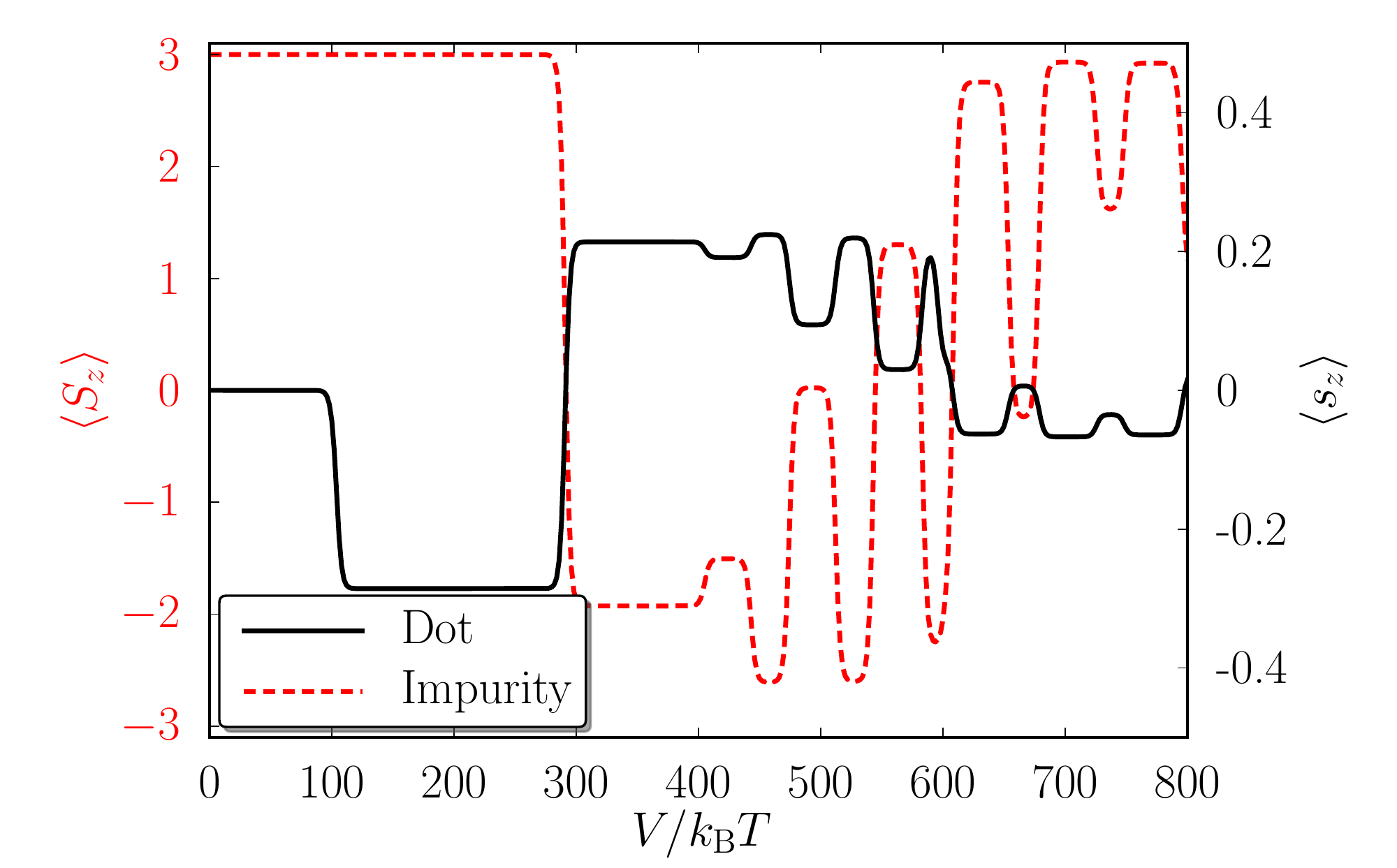}
	\caption{\label{fig:spin}(Color online) Average $z$ component of dot and impurity spin as a function of bias voltage for $\eta=+1$. Other parameters are the same as in Fig. \ref{fig:current}.}
\end{figure}
We now turn to our main result for the model containing an impurity spin in the barrier, that is we discuss the interplay between the dot and impurity spin and its manifestation in the transport properties. In particular, we investigate how the the dynamics of the dot and impurity spin gives rise to the sequence of positive and negative differential conductance features at the red and yellow dotted lines in Fig.~\ref{fig:conductance} for polarized electrodes and which is manifest in the current oscillations of the $I-V$ characteristics shown in Fig.~\ref{fig:current}. (In our discussion, we focus on the case $B>0$. For $B<0$, the impurity ground state is $\ket{-S}$ and basically the role of the cases $\eta=+1$ and $\eta=-1$ are interchanged.)

As tunneling into a ferromagnet is spin dependent, we find that the rates for the impurity spin transitions $\ket{m}\to\ket{m+1}$ and $\ket{m}\to\ket{m-1}$ are different, in general, because one transition involves tunneling of a minority spin electron, while the other involves tunneling of a majority spin electron. We therefore have for the rates $W_{m\to m+1}\propto 1-p$ and $W_{m\to m-1}\propto 1+p$. As a consequence, once exciting the impurity spin becomes energetically possible, the spin-polarized current through the right barrier has the tendency to flip the impurity spin into the state $\ket{-S}$ which in Figs.~\ref{fig:current} and~\ref{fig:spin} occurs at $V=2(\varepsilon+U+\Delta)=295\kB T$.

At larger bias, transitions where energy is gained from the impurity come into play. Alternatingly, they either lower the $z$ component of the impurity spin, $\ket{-S+i}\to\ket{-S+i-1}$, or raise it, $\ket{S-i}\to\ket{S-i+1}$, with $i\in[0,S]$. Hence, the expectation value of $S_z$ is found to oscillate as a function of bias voltage as is shown in Fig.~\ref{fig:spin}.

The other key ingredient for understanding the conductance oscillations is the fact that tunneling through the right barrier is spin dependent in two respects. On the one hand, there is the dependence on the spin of the tunneling electron due to the polarization of the lead, mentioned already above. However, the tunneling also depends on the state of the impurity spin as the tunneling rate for spin-up (down) electrons is proportional to $|t_\text{R}+S_z j_\text{R}|^2$ ($|t_\text{R}-S_z j_\text{R}|^2$). This kind of spin dependence is, then, responsible for relating the dot spin to the impurity spin state. If, e.g., the impurity is in the ground state $\ket{+S}$, spin up electrons can leave more easily to the right lead than spin down electrons. This results in a spin down accumulating on the quantum dot, see Fig.~\ref{fig:spin}, $2(\varepsilon+U)<V<2(\varepsilon+U+\Delta)$. When the $z$ component of the impurity spin has a negative expectation value, the situation is reversed. Now, spin down leaves the dot more easily such that spin up accumulates on the dot.

The spin accumulation on the dot affects the current through the system.
If the lower dot level is occupied by a spin up electron, the Pauli principle prevents a second spin up electron to enter the dot while transport of spin down electrons is suppressed by their smaller density of states. Hence, the accumulation of spin up on the dot suppresses the current. On the contrary, a spin down in the lower level does hardly affect the current, as it can proceed by spin up electrons tunneling through the upper level.

Hence, in symmary we find that the interplay between the current-induced switching of the impurity spin and the spin-dependent tunneling through the right barrier results in an interesting spin dynamics in the quantum-dot spin valve which manifests itself in the transport properties of the system. We emphasize that the transport signatures discussed here are present for moderate polarizations. Hence, the polarizations of Fe, Co, or Ni (Refs.~\onlinecite{Monsma_spin_2000,pasupathy_kondo_2004}) should be sufficient to experimentally detect them. For very large polarizations, the current oscillations discussed above are even absent, because in this case the current-induced switching mechanism is so strong that the impurity will be kept in state $\ket{-S}$ once spin excitation becomes possible.

\subsubsection{\label{sssec:fano}Giant Fano factor}
\begin{figure}
	\includegraphics[width=\columnwidth]{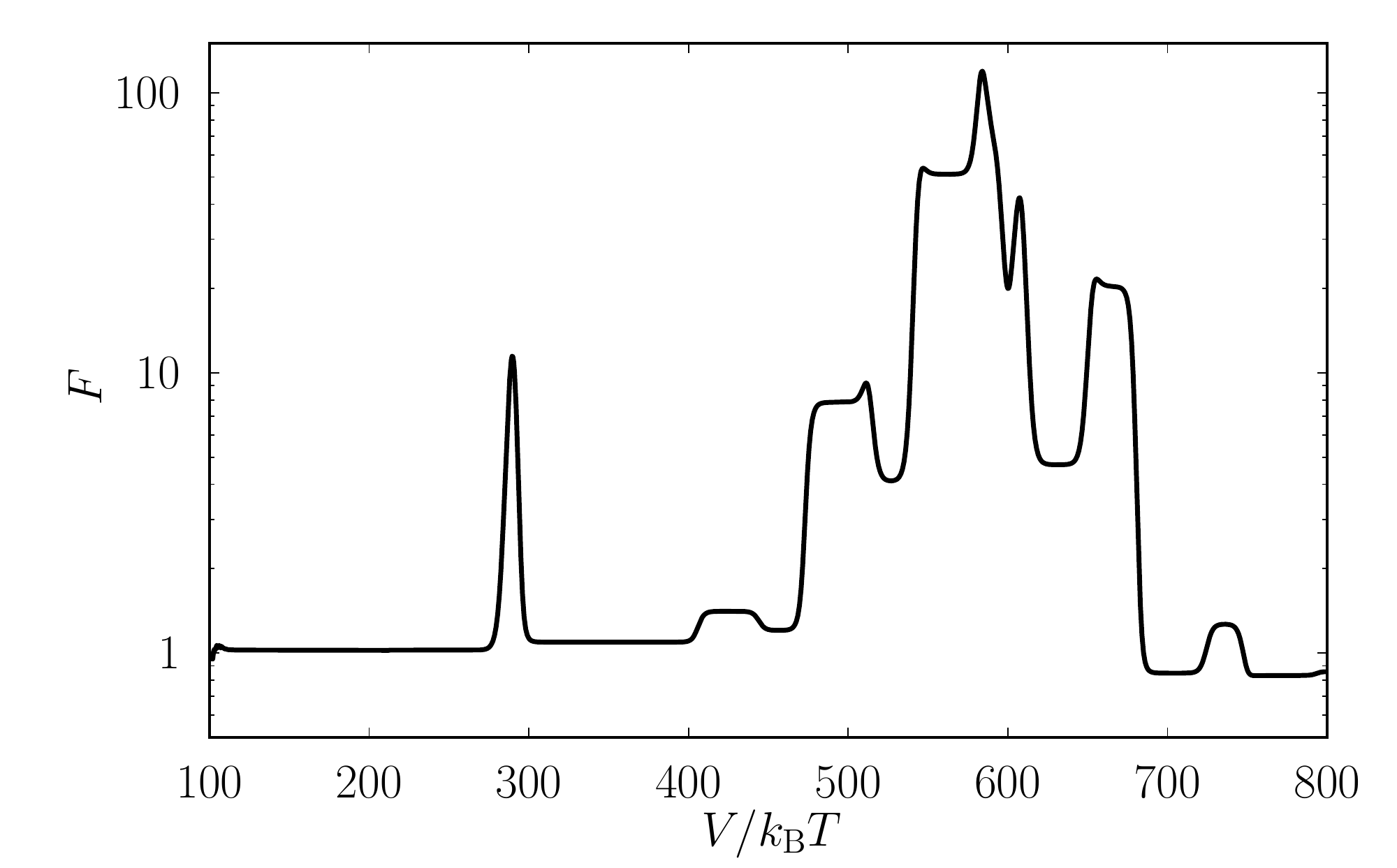}
	\caption{\label{fig:Fano}(Color online) Fano factor as function of bias voltage. Parameters are the same as in Fig.~\ref{fig:spin}. The sharp peak at $V=295\kB T$ arises at the onset of spin excitations (dashed, blue line in Fig.~\ref{fig:conductance}).}
\end{figure}
\begin{figure}
	\includegraphics[width=\columnwidth]{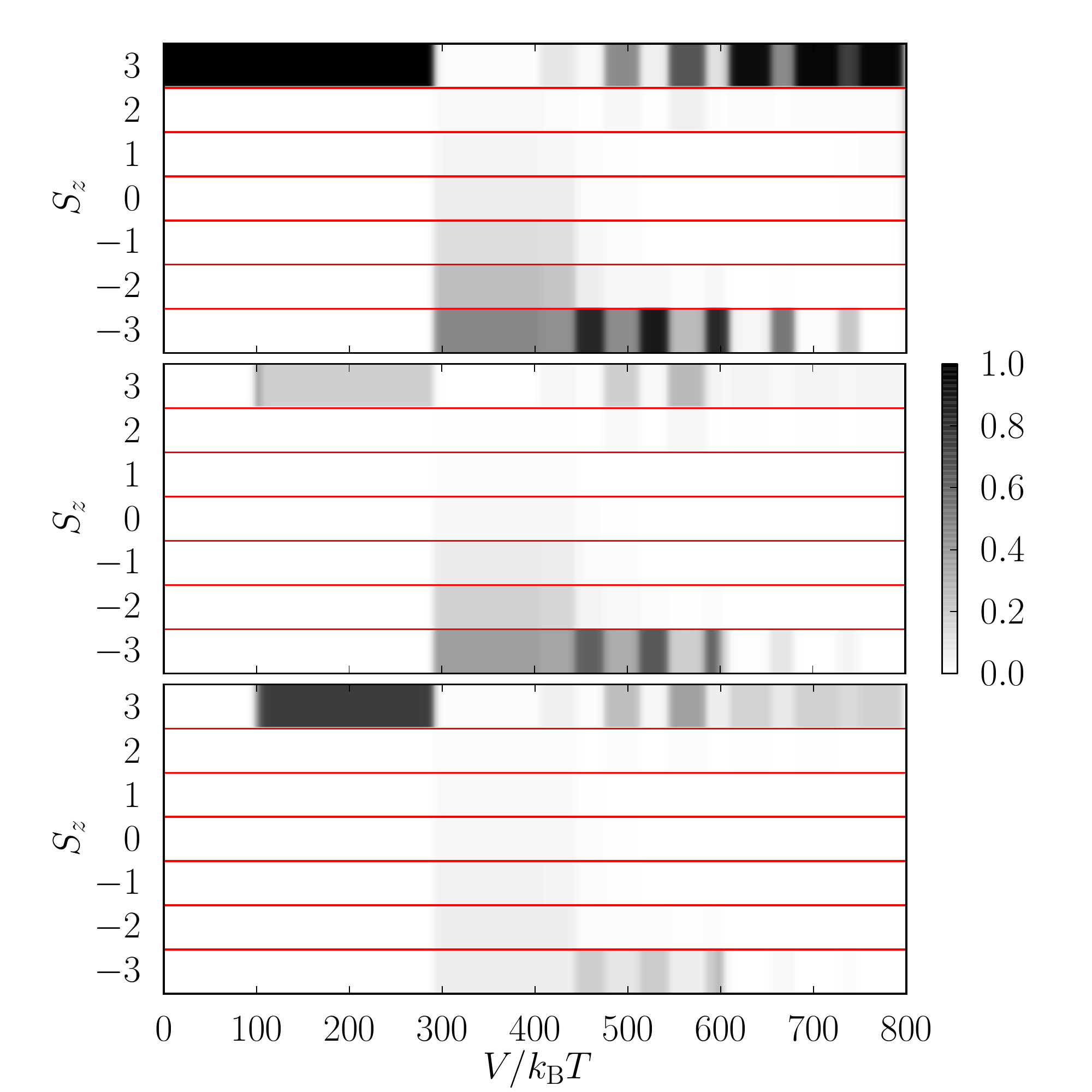}
	\caption{\label{fig:spinprob}Probabilities to find the impurity in spin state $\ket{S_z}$ as a function of bias voltage (upper panel). The middle (lower) panel shows the probability to find the system in the state $\ket{\up}\otimes\ket{S_z}$ ($\ket{\down}\otimes\ket{S_z}$). Parameters are the same as in Fig.~\ref{fig:spin}.}
\end{figure}
In the above discussion of the current oscillations, we found the current to be sensitive to the average impurity spin only. The current noise however sheds more light on the dynamics of the impurity spin. As can be seen in  Fig.~\ref{fig:Fano}, in the region where energy can be gained from the impurity to allow electrons to leave the dot to the right (corresponding to the region of dotted lines in Fig.~\ref{fig:conductance}), the Fano factor is strongly enhanced (note the logarithmic scale).

The mechanism which gives rise to these giant Fano factors is the following. As can be seen in the upper panel of Fig.~\ref{fig:spinprob}, the probability to find the impurity in the states $\ket{\pm S}$ is finite for both states in the region where the Fano factor becomes large. The dot spin follows the behavior of the impurity spin which means that it will point either up or down, depending on the sign of $S_z$ as shown in the middle and lower panel of Fig.~\ref{fig:spinprob}. The two dot spin configurations carry a different current, because only for the spin up accumulation transport through the dot becomes spin-blockaded. Hence, the system switches between two current states on a rather large time scale (as it takes several spin flip processes to reverse the impurity spin) which gives rise to random telegraph noise. Similar behavior can be found, e.g., in arrays of moveable colloid particles,~\cite{nishiguchi_shot-noise-induced_2002} transport through molecules with strong electron-phonon coupling and strong vibrational relaxation~\cite{koch_theory_2006} or double dot Aharonov-Bohm interferometers.~\cite{urban_tunable_2009}

It is interesting to note that the giant Fano factor occurs for parallel as well as antiparallel magnetizations. This is due to the fact that the moderate polarizations chosen here the effects of bunching for parallel magnetizations and spin blockade for antiparallel magnetizations are rather weak such that they are unimportant compared to the random telegraph switching.

\subsection{\label{ssec:noncoll}Noncollinear magnetizations}
So far, we only discussed the transport properties for nonmagnetic or collinearly magnetized electrodes. When dealing with noncollinear magnetizations, a new physical effect comes into play: the precession of the dot spin in an exchange field generated by virtual tunneling between the dot and the leads. As mentioned in the introduction, the interplay between spin accumulation and spin precession gives rise to a number of characteristic transport signatures. In the following, we discuss how the dependence of the exchange field on the state of the impurity spin influences the transport properties, in particular, the current and the finite-frequency Fano factor.

\subsubsection{\label{sssec:current}Current}
\begin{figure}
	\subfigure[]{\label{fig:currentperpV+}
	\includegraphics[width=\columnwidth]{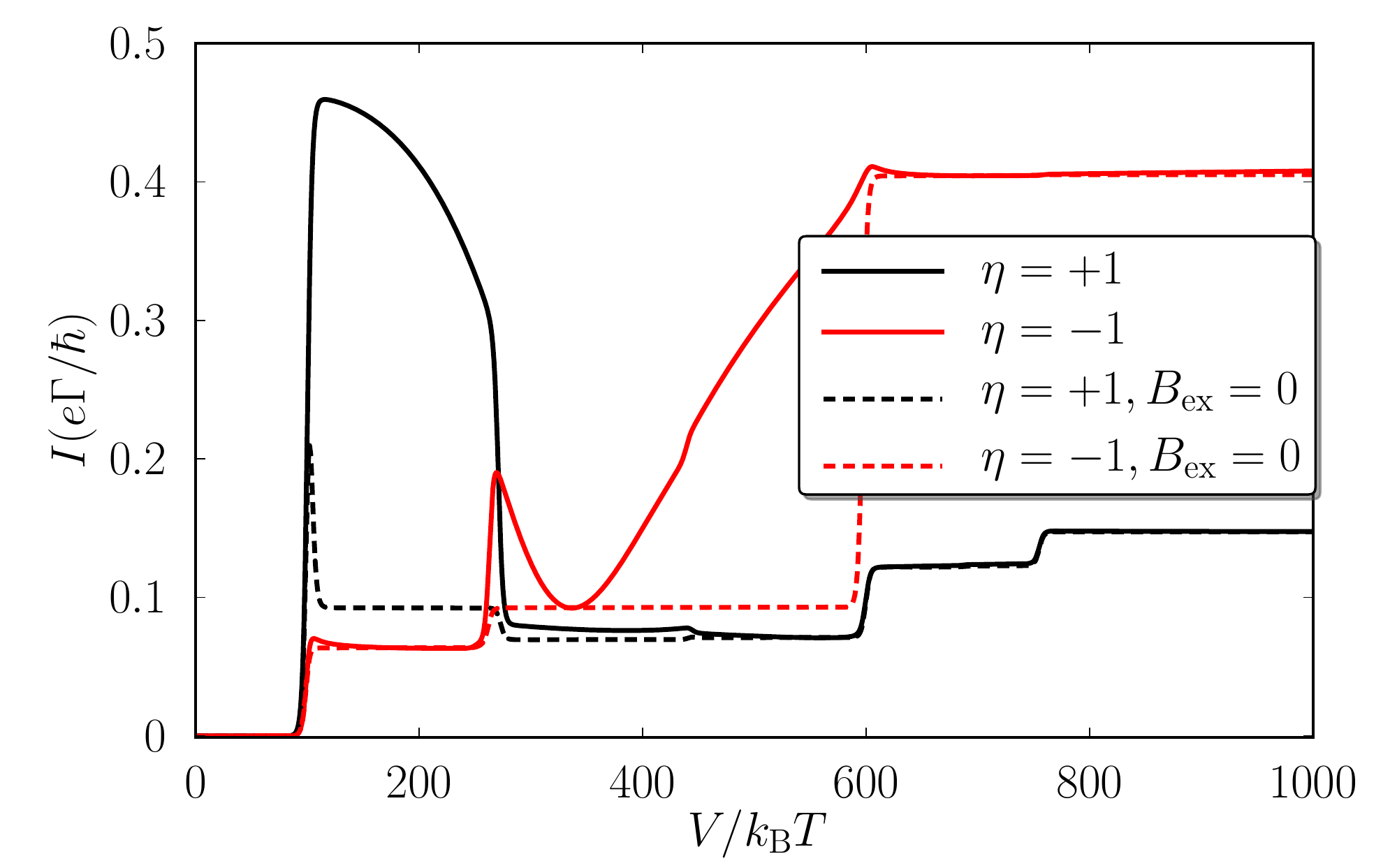}
	}
	\subfigure[]{\label{fig:currentperpV-}
	\includegraphics[width=\columnwidth]{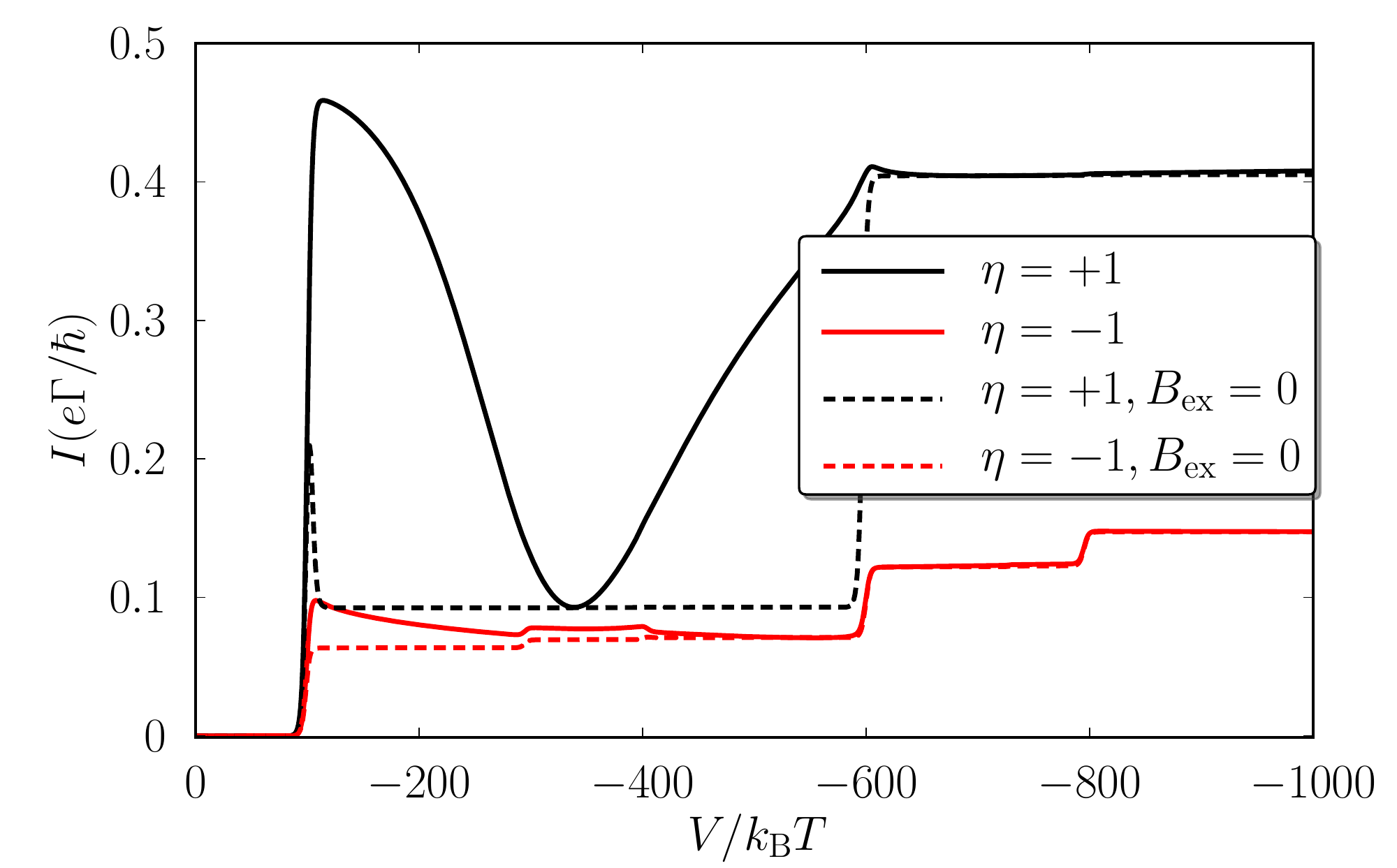}
	}
	\caption{\label{fig:currentperp}(Color online) Current as a function of bias voltage for perpendicular magnetizations for $p=0.9$ and $\Gamma_\text{L}=\Gamma_\text{R}=10J_\text{R}$. Black curves are for $\eta=+1$ while red curves are for $\eta=-1$. For the upper panel, $\varepsilon=-300\kB T$ while for the lower panel $\varepsilon=50\kB T$. Other parameters  as in Fig.~\ref{fig:Bexchange}. While the solid lines represent the full result, for the dashed curves the exchange field was set to zero by hand.}
\end{figure}
\begin{figure}
	\includegraphics[width=\columnwidth]{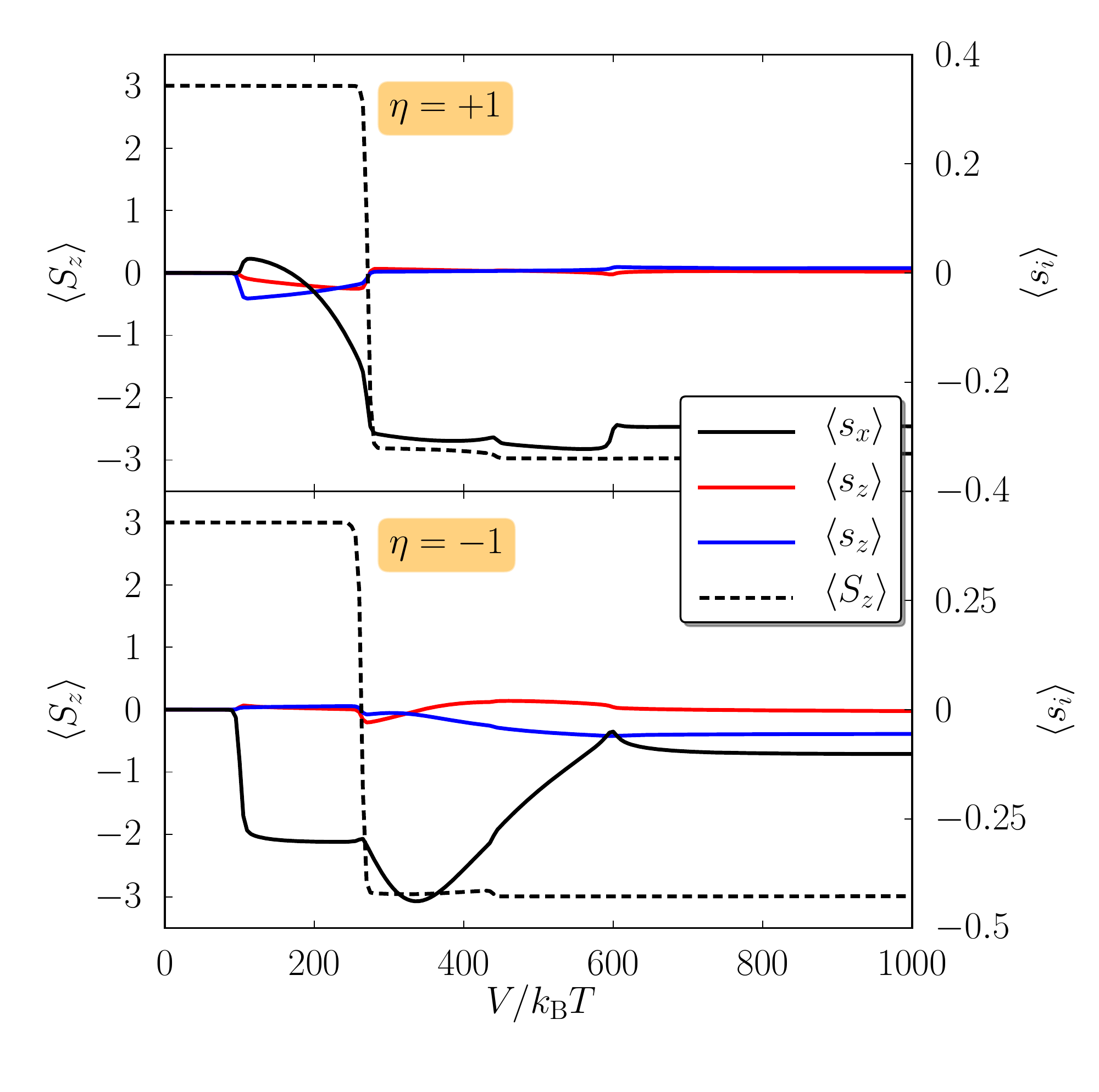}
	\caption{\label{fig:spinperp}(Color online) $x$, $y$ and $z$ component of the average dot spin and $z$ component of the average impurity spin for perpendicular magnetizations as function of bias voltage for positive (upper panel) and negative (lower panel) sign of the interference term. Parameters as in Fig.~\ref{fig:currentperpV+}.}
\end{figure}
In the normal quantum-dot spin valve, the current-voltage characteristics provides information about the exchange field acting on the dot spin. At the particle-hole symmetric point of the Anderson model ($\varepsilon=-U/2$), the exchange field vanishes. Hence, only the spin accumulation on the dot influences the current, leading to a reduced current in the noncollinear geometry compared to the parallel case. Away from this point, a finite exchange field acts on the dot spin and gives rise to a precession of the spin out of the blocking position. As a consequence, the current flowing through the system becomes larger than at the particle-hole symmetric point, resulting in a characteristic U-shaped $I-V$ curve with a broad region of negative differential conductance.

For the quantum-dot spin valve with an impurity embedded in one of the tunnel barriers studying the current as a function of bias voltage allows us to investigate the dependence of the exchange field on the state of the impurity. We will first focus on the situation where the interference term has positive sign and $V>0$. In this case, if the impurity is in state $\ket{+S}$, there is a large exchange field acting on the dot spin (cf. Fig.~\ref{fig:Bexchange}), giving rise to a strong precession and thereby to a clear lifting of the spin blockade. If the impurity is in state $\ket{-S}$, however, there is only a small exchange field acting on the dot. Consequently, the spin blockade persists and the current is not much enhanced by the exchange field.

As is shown in Fig.~\ref{fig:currentperpV+} (solid black curve), we therefore find a large current at the onset of transport through the quantum dot. When increasing the bias voltage, the current decreases slightly as the exchange field becomes weaker. When the bias voltage is increased above the threshold for impurity excitations, the impurity is switched from its ground state $\ket{+S}$ into the state $\ket{-S}$ by the spin-polarized current (cf. upper panel of Fig.~\ref{fig:spinperp}). As a consequence, the exchange field changes abruptly at threshold towards smaller values. This implies that the spin blockade on the dot cannot be lifted anymore and the current is suppressed compared to its values below threshold.

When the interference term has a negative sign [red curves in Fig.~\ref{fig:currentperpV+}], the situation is reversed. Now the exchange field takes on small values when the impurity is in state $\ket{+S}$ while it takes large values when the impurity is in state $\ket{-S}$. Hence, the current is now suppressed below threshold while above threshold we find a nontrivial bias dependence of the current due to the energy dependence of the exchange field.

When a negative bias voltage is applied, no switching occurs at the impurity excitation threshold as the spin-polarized current has the tendency to bring the impurity into the state $\ket{+S}$ which is the ground state. We therefore find that now the exchange field effects are clearly visible for a positive sign of the interference term, while they are very small for a negative interference term.

In summary, we have shown how the current-voltage characteristics for noncollinear magnetizations provide access to the dependence of the exchange field on the impurity spin state as well as to the sign of the interference term. We note, however, that the observation of exchange field effects in the current relies on large polarizations.

\subsubsection{\label{sssec:finfreqnoise}Frequency-dependent Fano factor}
\begin{figure}
	\includegraphics[width=\columnwidth]{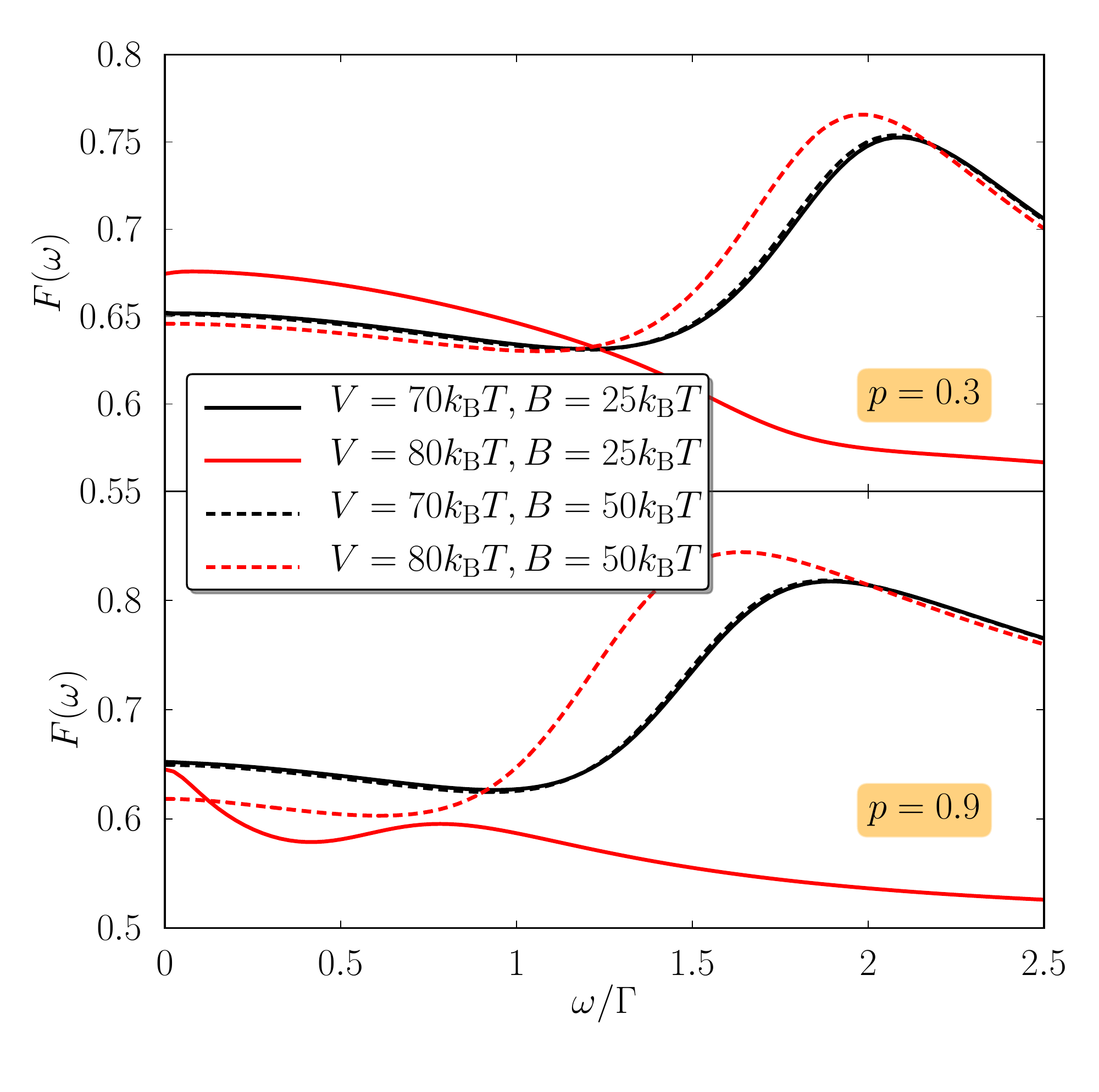}
	\caption{\label{fig:finfreqFano}(Color online) Frequency-dependent Fano factor for moderate ($p=0.3$, upper panel) and large ($p=0.9$, lower panel) polarizations. For $p=0.3$, an external magnetic field perpendicular $B_\text{ext}=1.5\Gamma$ is applied to the quantum dot alongs its quantization axis. Other parameters are the same as in Fig.~\ref{fig:currentperp}, except that now $D=0$ and $\eta=+1$.}
\end{figure}
The effects of the exchange field switching on the $I-V$ characteristic occur only for large polarizations as otherwise the spin blockade on the dot is too weak. Another method to gain information about the exchange field which also works for smaller polarizations is to study the frequency-dependent Fano factor.~\cite{braun_frequency-dependent_2006}

In Fig.~\ref{fig:finfreqFano}, the frequency-dependent Fano factor is shown for different bias voltages and different parameters of the spin Hamiltonian. Parameters are chosen such that for the smaller of the two bias voltages (black curves) the impurity spin cannot be excited. For the larger bias voltage, the impurity parameters allow an excitation of the impurity spin only for the red solid curves while for the red dashed curves the impurity still stays in the ground state.

For large polarizations and voltages below the excitation threshold, the finite-frequency noise shows a peak at the Larmor frequency of the exchange field. As the impurity is in the ground state, the system is not sensitive to the impurity parameters and both curves practically coincide. If the bias is increased but the impurity stays in the ground state, the absolute value of the exchange field is slightly reduced, cf. lower panel of Fig.~\ref{fig:Bexchange}, resulting in a small shift of the resonance signal toward smaller frequencies. If the bias is increased and a switching of the impurity spin occurs, we instead find that the exchange field is significantly reduced, cf. Fig.~\ref{fig:Bexchange}, and therefore the resonance peak is also shifted to much smaller frequencies. Hence, by detecting the resonance frequency as a function of bias voltage one can gain information about the switching of the exchange field as a consequnce of the switching of the impurity spin.

For small polarizations, the effect does not work the same way as just described. Now, the exchange field just changes sign upon switching the impurity spin, cf. Fig.~\ref{fig:Bexchange}. Therefore, no clear shifting of the resonance position occurs. To overcome this problem, an external magnetic field of strength comparable to the exchange field can be applied perpendicularly to the plane defined by the electrode magnetizations. Now, the position is defined by the Larmor frequency of the total magnetic field while the form of the resonance signal depends on the relative angle between external field and exchange field as can be seen in the upper panel of Fig.~\ref{fig:finfreqFano}. If the impurity spin stays in the ground state, a peak shows up in $F(\omega)$ while a shoulder occurs at the Larmor frequency if the impurity spin can be switched by the current.
Hence, we have shown that the switching of the exchange field can be monitored also for moderate polarizations by measuring the finite-frequency current noise for a series of applied bias voltages.

\section{\label{sec:Dot}Results - Small spin on the dot}
In this section, we discuss the transport properties of a quantum-dot spin valve containing an additional spin $1/2$ on the quantum dot. We will focus on the two transport regimes (ii) and (v) introduced above, cf. Table~\ref{tab:superpositions}. In these regimes, the exchange coupling is small, $J\lesssim\Gamma$, making them particularly suited to describe the influence of nuclear spins on transport through a quantum-dot spin valve. In regime (ii), the externally applied magnetic field $B$ is much larger than the tunnel coupling between dot and leads. In this regime, we show that the coherent superpositions of the singlet and one of the triplet states do not show up in the current. However, they do lead to a signature in the finite-frequency Fano factor. In regime (v), the external magnetic field is weak, $B\lesssim\Gamma$. In this case, we discuss how to extract the exchange coupling and external field from measurements of the finite-frequency Fano factor.

\subsection{Large magnetic field}
\begin{figure}
	\includegraphics[width=.45\textwidth]{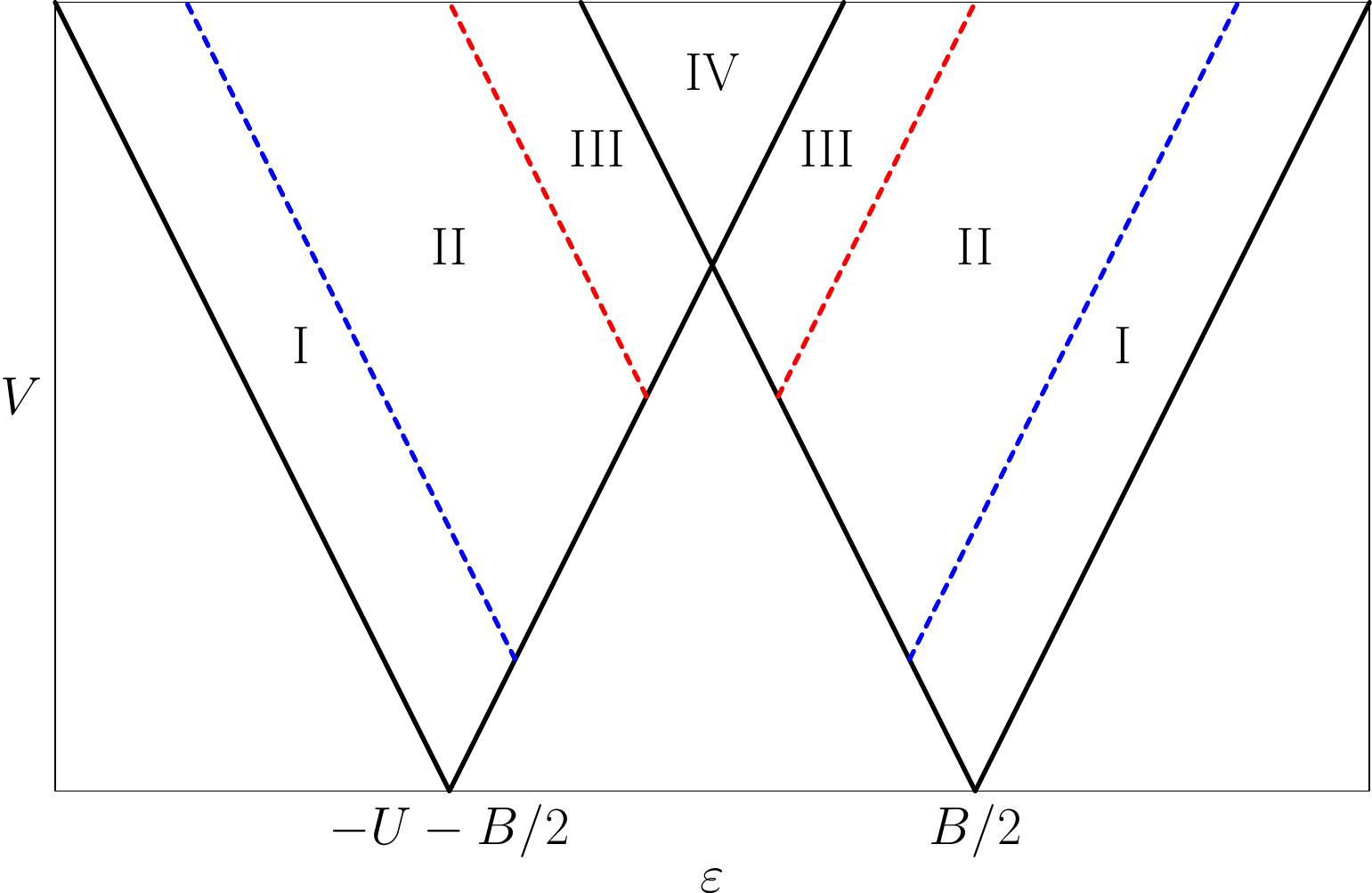}
	\caption{\label{fig:ConductanceScheme}(Color online) Schematic of the conductance for the situation where $B\gg\Gamma$, $J\lesssim\Gamma$. The lines indicate peaks in the differential conductance.}
\end{figure}
We start our discussion with the case $B\gg\Gamma$, $J\lesssim\Gamma$, where only superpositions of the singlet $S$ and the triplet state $T^0$ are relevant. In Fig.~\ref{fig:ConductanceScheme} we show schematically the differential conductance as a function of bias voltage $V$ and level position $\varepsilon$. The thick black lines indicate where a dot level is in resonance with either the left or right Fermi energy. Hence, they separate the Coulomb-blockaded regions from the regions where sequential through the dot is possible. For concreteness, we will now discuss the sequence of transport processes that come into play upon increasing the bias voltage for a fixed level position $\varepsilon=B/2$. For small bias voltages, transport is possible only through the states $\ket{0\down}$ and $\ket{T^-}$ as transitions from $\ket{0\down}$ to any of the other singly occupied states are energetically forbidden while transitions from $\ket{T^-}$ to $\ket{0\up}$ are impossible as they violate spin conservation.

When the bias voltage is increased across the blue dashed line, the bias window becomes large enough to also allow transitions from $\ket{0\down}$ to $\ket{T^0}$ and $\ket{S}$. As for both of these states there is a finite probability to find the impurity spin in state $\up$, both, $\ket{T^0}$ and $\ket{S}$, can serve as a starting point for a transition into the state $\ket{0\up}$ which then allows transitions into state $\ket{T^+}$. Hence, we find that as soon as the bias is large enough to excite the $\ket{0\down}$-$\ket{T^0}$ and $\ket{0\down}$-$\ket{S}$ transition, all empty and singly occupied dot states contribute to transport.

Upon increasing the bias voltage further across the red dashed line, another set of transport processes becomes energetically possible. While transitions from singly occupied dot state with the lowest energy, $\ket{T^-}$, to the doubly occupied state $\ket{d\down}$ are still not possible, it is nevertheless possible to occupy the dot with two electrons by taking as a starting point either $\ket{T^0}$ or $\ket{S}$.

Finally, when crossing another black line, the bias voltage is large enough to allow transitions between any two dot states that conserve spin.

Approximating the Fermi function as step functions, one can derive the following analytical expressions for the current through the quantum dot in the different transport regimes:
\begin{align}
	I_\text{I}&=\frac{\Gamma_\text{L}\Gamma_\text{R}}{\Gamma_\text{L}+\Gamma_\text{R}}=\frac{1-a^2}{2}\Gamma,\\
	I_\text{II}&=\frac{2\Gamma_\text{L}\Gamma_\text{R}}{2\Gamma_\text{L}+\Gamma_\text{R}}=\frac{2(1-a^2)}{3+a}\Gamma,\\
	I_\text{III}&=\frac{\Gamma_\text{L}\Gamma_\text{R}(\Gamma_\text{L}+2\Gamma_\text{R})}{\Gamma_\text{L}+\Gamma_\text{R}}=\frac{(3-a)(1-a^2)}{4}\Gamma,\\
	I_\text{IV}&=\frac{2\Gamma_\text{L}\Gamma_\text{R}}{\Gamma_\text{L}+\Gamma_\text{R}}=(1-a^2)\Gamma.
\end{align}
The Fano factor at zero frequency is given by
\begin{align}
	F_\text{I}&=\frac{\Gamma_\text{L}^2+\Gamma_\text{R}^2}{(\Gamma_\text{L}+\Gamma_\text{R})^2}=\frac{1}{2}(1+a^2),\\
	F_\text{II}&=\frac{4\Gamma_\text{L}^2+\Gamma_\text{R}^2}{(2\Gamma_\text{L}+\Gamma_\text{R})^2}=\frac{5+6a+5a^2}{(3+a)^2},\\
	F_\text{III}&=\frac{\Gamma_\text{L}^3+3\Gamma_\text{L}^2\Gamma_\text{R}+\Gamma_\text{R}^2}{(\Gamma_\text{L}+\Gamma_\text{R})^3}=\frac{1}{8}(5+3a+3a^2-3a^3),\\
	F_\text{IV}&=\frac{\Gamma_\text{L}^2+\Gamma_\text{R}^2}{(\Gamma_\text{L}+\Gamma_\text{R})^2}=\frac{1}{2}(1+a^2).
\end{align}
It is interesting to note that these expressions are precisely the same as found by Thielmann et al.~\cite{thielmann_shot_2003} for transport through a single-level quantum dot subject to a large magnetic field coupled to normal leads. Hence, in the chosen parameter regime, the current through the system and the zero-frequency Fano factor are neither sensitive to the presence of the impurity spin nor to the presence of ferromagnetic leads. The absence of any magnetoresistance is due to the presence of the large external field perpendicular to the plane spanned by the magnetizations which renders the system insensitive to the relative orientation of the magnetizations in this plane.

The fact that the system is insensitive to the presence of the impurity spin and to the coherent superpositions between $\ket{S}$ and $\ket{T^0}$ deserves some further investigation. Typically in systems where coherent superpositions of different states have to be taken into account as, e.g., a normal quantum-dot spin valve~\cite{kaenig_interaction-driven_2003,braun_theory_2004} or double quantum dots,~\cite{wunsch_probing_2005} the current shows a nontrivial bias dependence with broad regions of negative differential conductance due to the energy-dependent level renormalization that arises from virtual tunneling between the dot and the leads. Such effects are clearly absent here. To understand this behavior, we analyze the term describing the isospin accumulation, Eq.~\eqref{eq:SpinAccumulation}, in more detail. Due to the Fermi functions in this expression, a finite isospin accumulation can arise only in the regions marked III in Fig.~\ref{fig:ConductanceScheme}. When plugging in the expressions for the occupation probabilities, we find however that
\begin{multline}
	\left(\frac{d\vec I}{dt}\right)_\text{acc}=\left(-\Gamma_\text{L}\frac{\Gamma_\text{L}^2\Gamma_\text{R}^2}{4(\Gamma_\text{L}+\Gamma_\text{R})^2(\Gamma_\text{L}^2+2\Gamma_\text{L}\Gamma_\text{R}+2\Gamma_\text{R}^2)}\right.\\\left.
	+\Gamma_\text{R}\frac{\Gamma_\text{L}^3\Gamma_\text{R}}{4(\Gamma_\text{L}+\Gamma_\text{R})^2(\Gamma_\text{L}^2+2\Gamma_\text{L}\Gamma_\text{R}+2\Gamma_\text{R}^2)}\right)\vec e_x=0,
\end{multline}
i.e., the contributions to the isospin accumulation from the left and right lead cancel precisely. In consequence, the isospin vanishes on average in the stationary state such that the aforementioned level renormalization effects cannot influence the current and zero-frequency noise.

\begin{figure}
	\includegraphics[width=.45\textwidth]{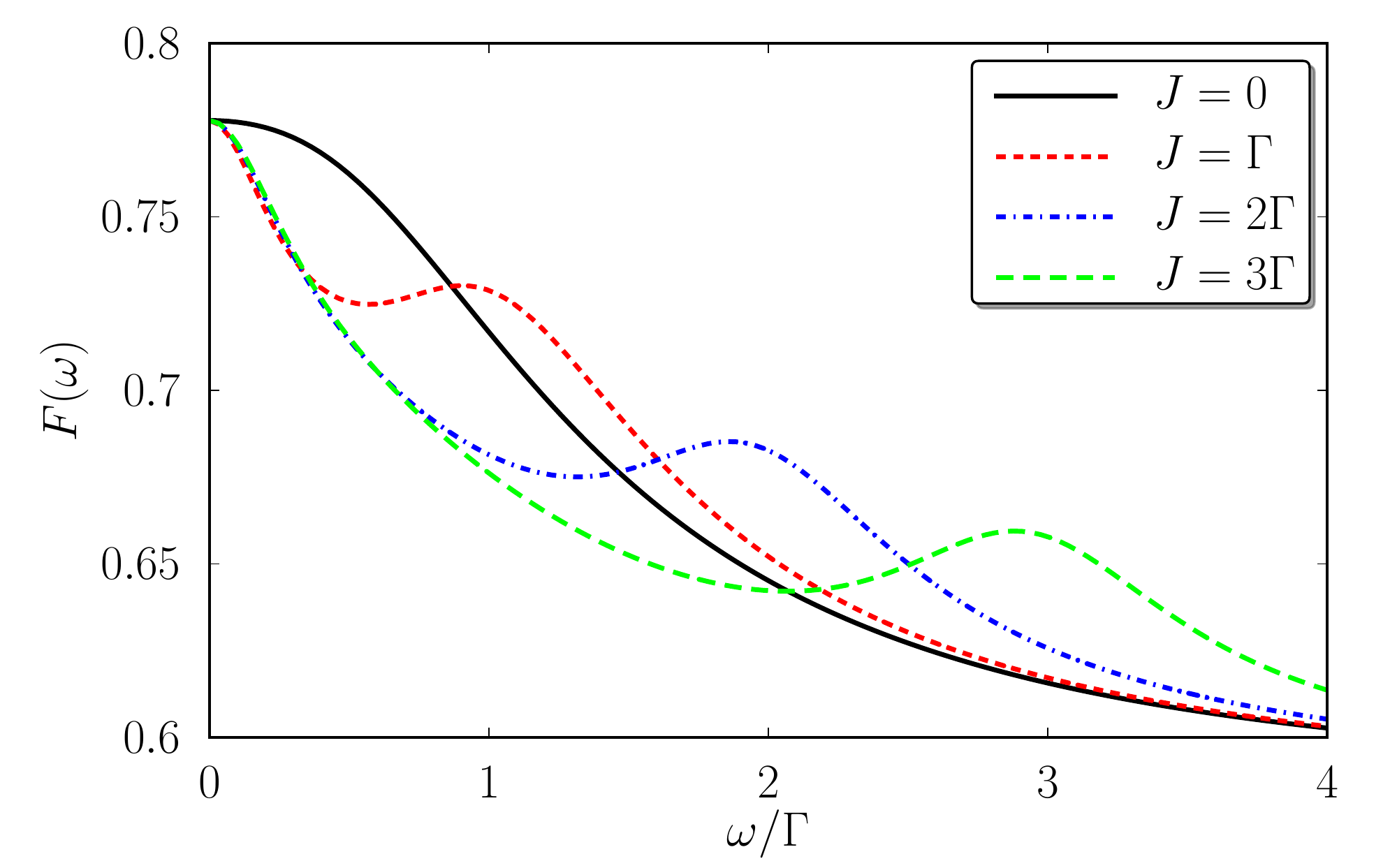}
	\caption{\label{fig:FinFreqNoise}(Color online) Finite-frequency Fano factor for different values of the exchange coupling $J$. The Fano factor shows a peak at the Larmor frequency associated with the exchange field. Parameters are $\Gamma_\text{L}=2\Gamma_\text{R}$, $\kB T=10\Gamma_\text{L}$, $B=50\kB T$, $U=5B$, $\varepsilon=B/2$ and $V=11B$.}
\end{figure}
So far, it seems that the coherent superpositions of $\ket{S}$ and $\ket{T^0}$ do not influence the transport through the system at all, such that a description in terms of ordinary rate equations which neglects the coherences suffices. This is not true, however, as can be seen from the finite-frequency Fano factor, which is shown in Fig.~\ref{fig:FinFreqNoise}. Here, apart from the usual peak which arises at zero frequency, the Fano factor additionally shows a peak associated with the Larmor frequency of the exchange field in Eq.~\eqref{eq:ExchangeField}. The coherent superpositions play a role here as the finite-frequency noise is sensitive to the dynamics of the dot spins while the current only captures stationary properties of the quantum dot. While the contributions form the left and right lead to the isospin accumulation cancel on average, indivdual processes can nevertheless give a finite isospin accumulation.

In contrast to the normal quantum-dot spin valve, where the Larmor frequency associated with the exchange field depends on the level position and applied bias voltage,~\cite{braun_frequency-dependent_2006} we find that in the system under investigation here, the Larmor frequency is simply given by the exchange coupling strength $J$. This is due to the fact that the energy-dependent contributions to the exchange field arise only in the $x$ component which is parallel to the isospin accumulating on the dot and hence cannot influence its precessional motion. The frequency-dependent Fano factor may, therefore, be used to experimentally determine the exchange coupling.

\subsection{Small magnetic field}
\begin{figure}
	\includegraphics[width=.5\textwidth]{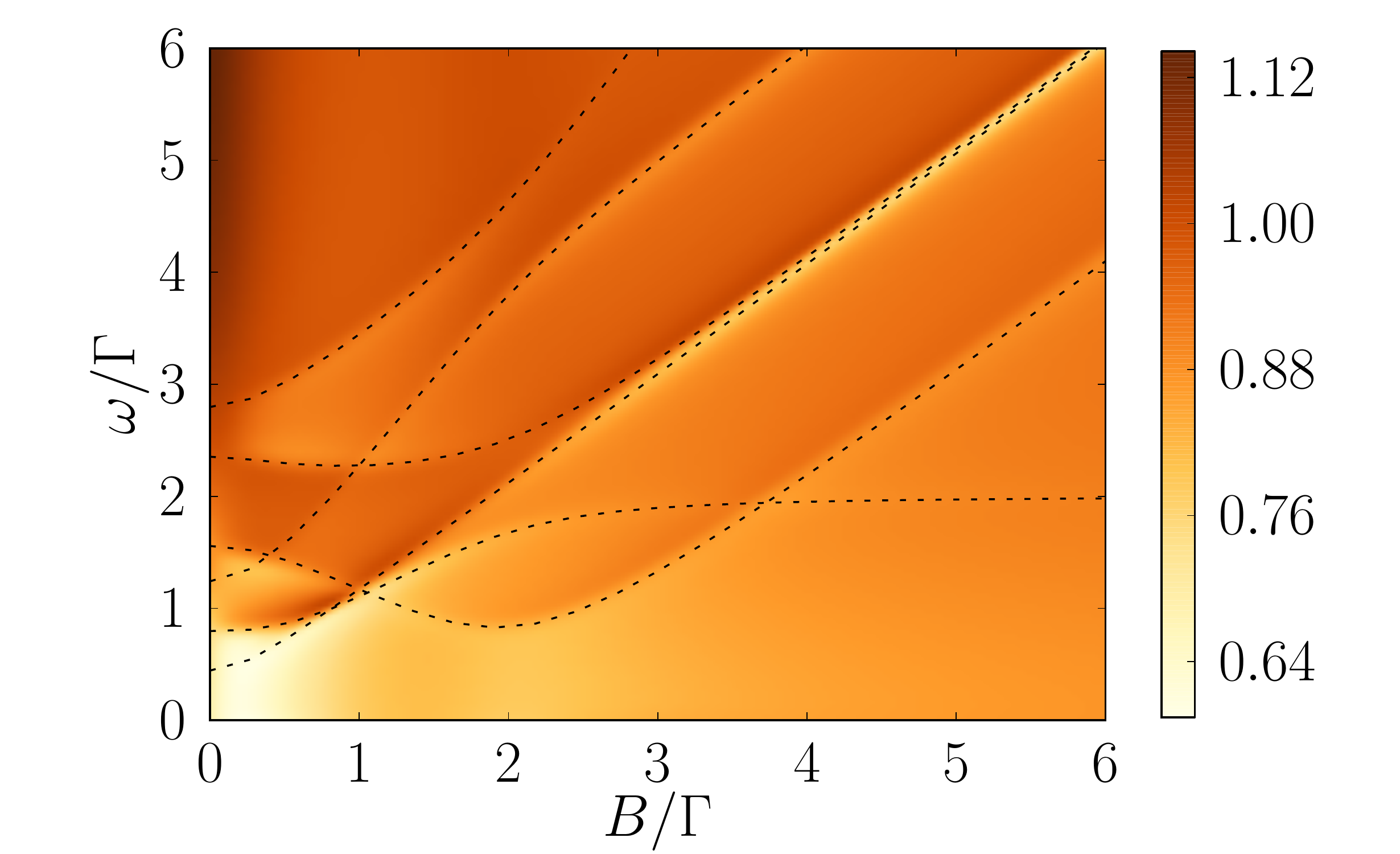}
	\includegraphics[width=.5\textwidth]{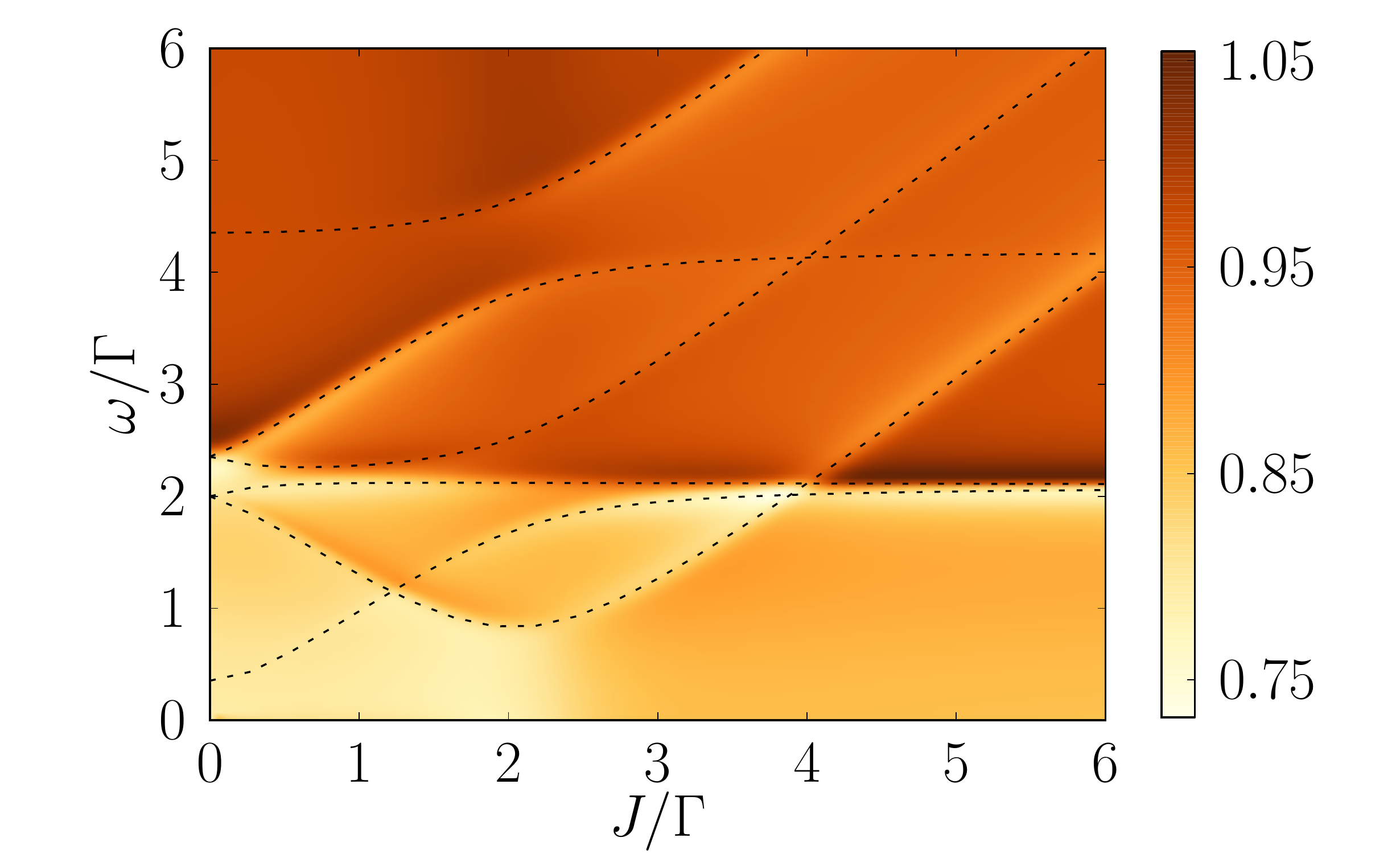}
	\caption{\label{fig:FinFreqNoiseII}(Color online) Frequency-dependent Fano factor as a function of frequency and external magnetic field (upper panel) or exchange coupling (lower panel). The level splittings calculated for a system of two exchange coupled spins in a magnetic field which is the sum of the exchange fields and the external fields are indicated by dotted lines. Parameters are $J=2\Gamma$ (upper panel), $B=2\Gamma$ (lower panel), $\Gamma_\text{L}=10\Gamma_\text{R}=0.05\kB T$,  $\varepsilon=50\kB T$, $V=105\kB T$, $U=150\kB T$, $p=0.9$, $\phi=\pi/2$.}
\end{figure}
In the last section, we saw that coherent superpositions can give rise to a signal in the frequency-dependent Fano factor $F(\omega)$ which could be interpreted as the precession of the isospin with the Larmor frequency of the exchange field. In the case where $B,J\lesssim\Gamma$, the situation is more complicated as now superpositions between any two states with the same number of electrons on the dot have to be taken into account.

In Fig.~\ref{fig:FinFreqNoiseII}, we show the finite-frequency Fano factor $F(\omega)$ as a function of the frequency $\omega$ and the externally applied magnetic field $B$ for fixed exchange coupling $J$ and vice versa. In both cases, the Fano factor shows a number of features at frequencies which all show a nontrivial dependence on the external field $B$ and the exchange coupling $J$. To gain a better understanding of these features, we consider a simpel spin model for the quantum dot. This is reasonable as the dot is singly occupied most of the time for the chosen parameters due to the asymmetric coupling to the leads. We model the dot as consisting of two spin $1/2$ particles that are exchange coupled and subject to external magnetic fields. While the impurity spin only couples to the externally applied field, the electron spin additionally experiences the exchange field generated by quantum charge fluctuations on the dot. Hence, the Hamiltonian of our model is given by
\begin{equation}\label{eq:SpinModel}
	H=J\vec S_1\cdot\vec S_2+\vec B_1\cdot \vec S_1+\vec B_2\cdot\vec S_2,
\end{equation}
where $\vec B_1=B_\text{ex,L}\vec n_\text{L}+B_\text{ex,R}\vec n_\text{R}+B\vec e_z$ is the sum of the external magnetic field and the exchange field while $\vec B_2=B\vec e_z$ is simply given by the external field. To show that this is indeed the correct way to describe the quantum dot system, in Appendix~\ref{app:2spinEOM} we give the equation of motion for the density matrix elements of the two spins. Comparing them to the master equation for the quantum dot system, Eqs.~\eqref{eq:occupations}-\eqref{eq:quadrupole}, shows the correspondence between the two systems. The only difference is the absence of dissipative terms in the equations for the spin model.

The differences between the energies of the various eigenstates of Eq.~\eqref{eq:SpinModel} as a function of exchange coupling and external field are indicated as black dotted lines in Fig.~\ref{fig:FinFreqNoiseII}. We find a nice agreement between these energy differences and the features observed in the Fano factor.
On the one hand, this indicates that the finite-frequency Fano factor is indeed sensitive to level splittings comparable to the tunnel couplings which cannot be resolved in the differential conductance as the conductance peaks are broadened by temperature which satisfies $\Gamma\ll\kB T$. This can provide experimental access, e.g., to the coupling between the dot and impurity spin.

On the other hand, our results nicely demonstrate that the exchange field which arises due to virtual tunneling processes and \emph{not} due to stray fields from the ferromagnetic leads, acts only on the electron spin but not on the impurity spin. If one could measure the Fano factor at finite frequency as computed above, one could therefore clearly distinguish exchange field effects from stray field effects as the latter ones would act in the same way on both spins and thereby give rise to a trivial dependence of the energy differences on the exchange coupling and the external field.

\section{\label{sec:Conclusion}Conclusions}
In this paper, we investigated transport through quantum-dot spin valves containing magnetic impurities. First, we considered the case where a large, anisotropic impurity spin is localized in one of the tunneling barrier. Our main focus was on the spin excitation and spin switching on the impurity spin. We pointed out how the Coulomb interaction on the dot allows a more detailed spectroscopy of the impurity spin. For magnetic electrodes, the spin-polarized current through the system can switch the spin of the impurity. As the state of the impurity spin influences the current, an interplay between the current and the impurity arises which gives rise to a series of positive and negative differential conductance as a function of bias voltage. We, furthermore, found that the dynamics of the impurity spin gives rise to a random telegraph signal associated with a huge Fano factor. Additionally, we found that for tunneling through the barrier containing the impurity interference between direct and exchange tunneling takes place even for nonmagnetic systems. This is in contrast to tunnel barriers where such interference effects arise only for magnetic electrodes. Finally, we found that the exchange field acting on the dot spin becomes dependent on the impurity spin state. This dependence allows the observation of the impurity switching in the current as well as in the finite-frequency noise for noncollinear magnetizations.

Second, we considered a spin $1/2$  impurity localized on the dot in the regime of small exchange coupling to the electron spin. Here, our main focus was on the coherent spin dynamics on the quantum dot. For a large external magnetic field, only coherent superpositions between the singlet $\ket{S}$ and triplet $\ket{T^0}$ are relevant. However, they influence neither the current nor the zero-frequency noise as they vanish in the stationary state. Nevertheless, as the superpositions can be excited, they give rise to a signal in the finite-frequency noise at $\omega=J$. For small external magnetic fields, all coherences have to be taken into account. They give rise to a large number of features in the finite-frequency noise. These provide informations about the level splittings that are influenced by renormalization effects due to virtual tunneling to the leads.

To summarize, we found that in both cases, for the large as well as for the small impurity spin, there is an intricate interplay between the electron and impurity spin that manifests itself in the transport properties of the system. The details of of this interplay vary, however.
The large impurity spin does not show any coherent dynamics in the sequential tunneling regime as its level splittings generated by the anisotropy are large compared to the tunnel couplings and the temperature. In consequence, these splittings can be resolved in the current which also allows to study the current-induced switching of the impurity. 
In contrast, for a small impurity spin, both, the electron and the impurity spin behave as one object that displays a coherent dynamics as level splittings are of the order of the tunnel coupling. Clear signatures of this dynamics can be found in the finite-frequency noise that allows to resolve level splittings of the order $\Gamma\ll\kB T$.

\acknowledgments
We acknowledge valuable discussion with Maarten Wegewijs, Sourin Das, Michael Baumgärtel, and Michael Hell. Financial support from DFG via SFB 491 is gratefully acknowledged.

\appendix
\section{\label{sec:Rules}Diagrammatic rules}
In this appendix, we summarize the diagrammatic rules to evaluate the self-energies entering the generalized master Eq.~\eqref{eq:MasterEquation} and the expressions for the current in Eq.~\eqref{eq:current} and current noise in Eq.~\eqref{eq:finfreqnoise}. To compute the self-energy $W_{\chi_2\chi_2'}^{\chi_1\chi_1'}$,
\begin{enumerate}
	\item Draw all topological different diagrams with tunneling lines connecting vertices on either the same or opposite propagators. Assign to the four corners and all propagators states $\chi$ and corresponding energies $E_\chi$ as well as an energy $\omega$ for every tunneling line.
	\item For each part of the diagram between adjacent vertices, assign a resolvent $1/(\Delta E+i0^+)$ where $\Delta E$ is the difference between the energies of left- and right-going tunneling lines and propagators.
	\item For each tunneling line, the diagram acquires a factor of $\frac{1}{2\pi}f_r^\pm(\omega)$, where the sign is determined by whether the line runs forward (-) or backward (+) with respect to the Keldysh contour.
	\item For each pair of vertices connected by a tunneling line the diagram is multiplied by $\frac{1+p}{2}\rho_r\bra{\chi_\text{i}'}C_{r\up}\ket{\chi_\text{i}}\bra{\chi_\text{f}'}C_{r\up}^\dagger\ket{\chi_\text{f}}+\frac{1-p}{2}\rho_r\bra{\chi_\text{i}'}C_{r\down}\ket{\chi_\text{i}}\bra{\chi_\text{f}'}C_{r\down}^\dagger\ket{\chi_\text{f}}$, where $\chi_\text{i}$ and $\chi_\text{i}'$ ($\chi_\text{f}$ and $\chi_\text{f}'$) are the states that enter and leave the vertex the tunneling line begins (ends) at, respectively. The operators $C_{r\up}$ and $C_{r\down}$ are the coefficients (including the dot operators!) of $a_{r\vec k+}^\dagger$ and $a_{r\vec k-}^\dagger$ in the tunnel Hamiltonians \eqref{eq:HtunL}, \eqref{eq:HtunR} and \eqref{eq:Htun}, respectively.
	\item Each vertex connecting the states $d$ and $\down$ gives rise to a factor of $-1$.
	\item The diagram obtains a factor $(-i)(-1)^{a+b}$, where $a$ is the number of vertices on the lower propagator and $b$ is the number of crossings of tunneling lines.
	\item Sum over all leads $r$ and integrate over all energies $\omega$.
\end{enumerate}
The diagrams for $W_{\phantom{I}\chi_2\chi_2'}^{I\chi_1\chi_1'}$ and $W_{\phantom{II}\chi_2\chi_2'}^{II\chi_1\chi_1'}$ are obtained by replacing one, respectively two tunneling vertices by current vertices. These give rise to a factor of $+1/2$ (-$1/2$) if the vertex is on the upper (lower) branch and corresponds to an electron tunneling into the right (left) lead or out off the left (right) lead.

\section{\label{sec:MasterEquation}Master equation}
\subsection{\label{sec:MasterEquationBarrier}Model A: Large spin in the barrier}
In this appendix, we give explicit expressions for the various quantities that occurred in the master equations for the occupation probabilities and the spin, Eqs.~\eqref{eq:masterprob} and \eqref{eq:masterspin} of model A, i.e., a quantum-dot spin valve with an impurity spin embedded in the right tunnel barrier. Introducing $\gamma=\Gamma_\text{R}+m^2J_\text{R}+2mp\eta\sqrt{\Gamma_\text{R}J_\text{R}}$, $\tilde \gamma=p\Gamma_\text{R}+m^2pJ_\text{R}+2m\eta\sqrt{\Gamma_\text{R}J_\text{R}}$, $\alpha_\pm=B+(2m\pm1)D$, and $A_\pm(m)=S(S+1)-m(m\pm1)$, we find for the coupling to the left lead
\begin{widetext}
\begin{equation}
	\vec W_\text{L}^{(0)}=\Gamma_\text{L}
	\left(
	\begin{array}{ccc}
	-2\fpL(\varepsilon) & \fmL(\varepsilon) & 0 \\
	2\fpL(\varepsilon) & -\fmL(\varepsilon)-\fpL(\varepsilon+U) & 2\fmL(\varepsilon+U) \\
	0 & \fpL(\varepsilon+U) & -\fmL(\varepsilon+U)
	\end{array}
	\right),
\end{equation}
\begin{equation}
	V_\text{L}^{(0)}=2p\Gamma_\text{L}
	\left(
	\begin{array}{c}
	\fmL(\varepsilon)\\
	-\fmL(\varepsilon)+\fpL(\varepsilon+U)\\
	-\fpL(\varepsilon+U)
	\end{array}
	\right),
\end{equation}
\begin{equation}
	\left(\frac{d\vec s_m}{dt}\right)_\text{acc,L}^{(0)}=p\Gamma_\text{L}\left(\fpL(\varepsilon)P_{0,m}-\frac{\fmL(\epsilon)-\fpL(\varepsilon+U)}{2}P_{1,m}-\fmL(\varepsilon+U)P_{d,m}\right)\vec n_\text{L},
\end{equation}
\begin{equation}
	\left(\frac{d\vec s_m}{dt}\right)_\text{rel,L}^{(0)}=-\Gamma_\text{L}\left(\fmL(\varepsilon)+\fpL(\varepsilon+U)\right)\vec s_m,
\end{equation}
which is identical to the expressions found for the ordinary quantum-dot spin valve in Ref.~\onlinecite{braun_theory_2004}. For the coupling to the right lead, we have
\begin{equation}
	\vec W_\text{R}^{(0)}=
	\begin{pmatrix}
	W_{\text{R},00}^{(0)} & \gamma \fmR(\varepsilon) & 0 \\
	2\gamma \fpR(\varepsilon) & W_{\text{R},11}^{(0)} & 2\gamma \fmR(\varepsilon+U) \\
	0 & \gamma \fpR(\varepsilon+U) & W_{\text{R},dd}^{(0)} \\
	\end{pmatrix},
\end{equation}
where the diagonal matrix elements are determined by the sum rule $\vec e^T\vec W=0$,
\begin{equation}
	V_\text{R}^{(0)}=2
	\begin{pmatrix}
	\gamma \fmR(\varepsilon)\\
	\tilde\gamma\left[-\fmR(\varepsilon)+\fpR(\varepsilon+U)\right]+A_+(m-1)J_\text{R}\left[\frac{1+p}{2}\fmR(\varepsilon-\alpha_-)+\frac{1-p}{2}\fpR(\varepsilon+U+\alpha_-)\right]\\-A_-(m+1)J_\text{R}\left[\frac{1-p}{2}\fmR(\varepsilon+\alpha_+)+\frac{1+p}{2}\fpR(\varepsilon+U-\alpha_+]\right)\\
	\gamma \fpR(\varepsilon+U)
	\end{pmatrix},
\end{equation}
\begin{equation}
	\vec W_\text{R}^{(+1)}=A_-(m+1)J_\text{R}
	\begin{pmatrix}
	0 & \frac{1+p}{2}\fmR(\varepsilon-\alpha_+) & 0 \\
	(1-p)\fpR(\varepsilon+\alpha_+) & 0 & (1+p)\fmR(\varepsilon+U-\alpha_+) \\
	0 & \frac{1-p}{2}\fpR(\varepsilon+U+\alpha_+) & 0
	\end{pmatrix},
\end{equation}
\begin{equation}
	V_\text{R}^{(+1)}=A_-(m+1)J_\text{R}
	\begin{pmatrix}
	-(1+p)\fmR(\varepsilon-\alpha_+)\\
	0\\
	-(1-p)\fpR(\varepsilon+U+\alpha_+)
	\end{pmatrix},
\end{equation}
\begin{equation}
	\vec W_\text{R}^{(-1)}=A_+(m-1)J_\text{R}
	\begin{pmatrix}
		0 & \frac{1-p}{2}\fmR(\varepsilon+\alpha_-) & 0 \\
		(1+p)\fpR(\varepsilon-\alpha_-) & 0 & (1-p)\fmR(\varepsilon+U+\alpha_-) \\
		0 & \frac{1+p}{2}\fpR(\varepsilon+U-\alpha_-) & 0
	\end{pmatrix},
\end{equation}
\begin{equation}
	V_\text{R}^{(-1)}=A_+(m-1)J_\text{R}
	\begin{pmatrix}
	(1-p)\fmR(\varepsilon+\alpha_-)\\
	0\\
	(1+p)\fpR(\varepsilon+U-\alpha_-)
	\end{pmatrix},
\end{equation}
\begin{multline}
	\left(\frac{d\vec s_m}{dt}\right)_\text{acc,R}^{(0)}=\left\{\tilde\gamma \fpR(\varepsilon)P_{0,m}-\tilde\gamma \fmR(\varepsilon+U)P_{d,m}+\left[\tilde\gamma\frac{-\fmR(\varepsilon)+\fpR(\varepsilon+U)}{2}\right.\right.\\\left.\left.
	+A_+(m-1)\frac{J_\text{R}}{2}\left(\frac{1+p}{2}\fmR(\varepsilon-\alpha_-)+\frac{1-p}{2}\fpR(\varepsilon+U+\alpha_-)\right)\right.\right.\\\left.\left.
	-A_-(m+1)\frac{J_\text{R}}{2}\left(\frac{1-p}{2}\fmR(\varepsilon+\alpha_+)+\frac{1+p}{2}\fpR(\varepsilon+U-\alpha_+)\right)\right]P_{1,m}\right\}\vec n_\text{R},
\end{multline}
\begin{equation}
	\left(\frac{d\vec s_m}{dt}\right)_\text{acc,R}^{(+1)}=A_-(m+1)J_\text{R}\left[(1-p)\fpR(\varepsilon+\alpha_+)P_{0,m+1}+(1+p)\fmR(\varepsilon+U-\alpha_+)P_{d,m+1}\vphantom{\frac{1}{2}}\right]\vec n_\text{R},
\end{equation}
\begin{equation}
	\left(\frac{d\vec s_m}{dt}\right)_\text{acc,R}^{(-1)}=-A_+(m-1)J_\text{R}\left[(1+p)\fpR(\varepsilon-\alpha_-)P_{0,m-1}+(1-p)\fmR(\varepsilon+U+\alpha_-)P_{d,m-1}\vphantom{\frac{1}{2}}\right]\vec n_\text{R},
\end{equation}
\begin{multline}
	\left(\frac{d\vec s_m}{dt}\right)_\text{rel,R}^{(0)}=-\left[\gamma(\fmR(\varepsilon)+\fpR(\varepsilon+U))
	+A_+(m-1)J_\text{R}\left(\frac{1+p}{2}\fmR(\varepsilon-\alpha_-)+\frac{1-p}{2}\fpR(\varepsilon+U+\alpha_-)\right)\right.\\\left.
	+A_-(m+1)J_\text{R}\left(\frac{1-p}{2}\fmR(\varepsilon+\alpha_+)+\frac{1+p}{2}\fpR(\varepsilon+U-\alpha_+)\right)\right]\vec s_{m}.
\end{multline}

\subsection{\label{sec:MasterEquationDot}Model B: Small spin on the dot}
The vector $\vec V$ occurring in the master Eq.~\eqref{eq:MEP} of model B, i.e., the quantum-dot spin valve containing a spin $1/2$ on the dot is given by
\begin{equation}
	\vec V=\sum_r\Gamma_r\left(
	\begin{array}{c}
		-f_r^-(\varepsilon-B/2) \\
		f_r^-(\varepsilon+B/2) \\
		0 \\
		f_r^-(\varepsilon-B/2)-f_r^-(\varepsilon+B/2)+f_r^+(\varepsilon+U+B/2)-f_r^+(\varepsilon+U-B/2) \\
		0 \\
		-f_r^+(\varepsilon+U+B/2) \\
		f_r^+(\varepsilon+U-B/2)
	\end{array}
	\right).
\end{equation}
\end{widetext}

\section{\label{app:densitymatrix}Density matrix elements}
In this section, we give the relation between the physical quantities introduced in Sec.~\ref{sssec:case5} and the density matrix elements in Eq.~\eqref{eq:densitymatrix}.
For the occupation probabilities of the dot we have
\begin{align}
	P_0&=P_{0\up}+P_{0\down},\\
	P_1&=P_{T^+}+P_{T^0}+P_{T^-}+P_S,\\
	P_d&=P_{d\up}+P_{d\down}.
\end{align}
For the expectation values of the impurity spin when the dot is empty, we get
\begin{align}
	S_{0x}&=\re P^{0\down}_{0\up},\\
	S_{0y}&=\im P^{0\down}_{0\up},\\
	S_{0z}&=\frac{P_{0\up}-P_{0\down}}{2}.
\end{align}
When the dot is singly occupied, we have
\begin{align}
	S_{2x}&=\frac{-\re P^S_{T^-}+\re P^S_{T^+}+\re P^{T^0}_{T^-}+\re P^{T^0}_{T^+}}{\sqrt{2}},\\
	S_{2y}&=\frac{\im P^S_{T^-}+\im P^S_{T^+}+\im P^{T^0}_{T^+}-\im P^{T^0}_{T^-}}{\sqrt{2}},\\
	S_{2z}&=\frac{P_{T^+}-P_{T^-}-2\re P^S_{T^0}}{2},
\end{align}
and for the doubly occupied dot, we find
\begin{align}
	S_{dx}&=\re P^{d\down}_{d\up},\\
	S_{dy}&=\im P^{d\down}_{d\up},\\
	S_{dz}&=\frac{P_{d\up}-P_{d\down}}{2}.
\end{align}
The expectation value of the electron spin on the dot can be expressed as
\begin{align}
	S_{1x}&=\frac{\re P^S_{T^-}-\re P^S_{T^+}+\re P^{T^0}_{T^-}+\re P^{T^0}_{T^+}}{\sqrt{2}},\\
	S_{1y}&=\frac{-\im P^S_{T^-}-\im P^S_{T^+}+\im P^{T^0}_{T^+}-\im P^{T^0}_{T^-}}{\sqrt{2}},\\
	S_{1z}&=\frac{P_{T^+}-P_{T^-}+2\re P^S_{T^0}}{2}.
\end{align}
For the scalar product of the electron and impurity spin, we find
\begin{equation}
	\vec S_1\cdot\vec S_2=\frac{P_{T^+}+P_{T^0}+P_{T^-}-3P_S}{4},
\end{equation}
while for their vector product, we get
\begin{align}
	(\vec S_1\times\vec S_2)_x&=\frac{\im P^S_{T^-}-\im P^S_{T^+}}{\sqrt{2}},\\
	(\vec S_1\times\vec S_2)_y&=\frac{\re P^S_{T^+}+\re P^S_{T^-}}{\sqrt{2}},\\
	(\vec S_1\times\vec S_2)_z&=\im P^S_{T^0}.
\end{align}
Finally, the quadrupole moments are given by
\begin{align}
	Q_{xx}&=\frac{2P_{T^0}-P_{T^-}-P_{T^+}+6\re P^{T^-}_{T^+}}{12},\\
	Q_{yy}&=\frac{2P_{T^0}-P_{T^-}-P_{T^+}-6\re P^{T^-}_{T^+}}{12},\\
	Q_{zz}&=\frac{P_{T^+}-2P_{T^0}+P_{T^-}}{6},\\
	Q_{xy}&=\frac{\im P^{T^-}_{T^+}}{2},\\
	Q_{xz}&=\frac{-\re P^{T^0}_{T^-}+\re P^{T^0}_{T^+}}{2\sqrt{2}},\\
	Q_{yz}&=\frac{\im P^{T^0}_{T^-}+\im P^{T^0}_{T^+}}{2\sqrt{2}}.
\end{align}

\section{\label{app:2spinEOM}Equation of motion for two spin $1/2$ particles}
The equations of motion of two exchange-coupled spins $\vec S_1$ and $\vec S_2$ subject to magnetic fields $\vec B_1$ and $\vec B_2$, respectively, as described by the Hamiltonian~\eqref{eq:SpinModel} are given by
\begin{equation}
	\frac{d\vec S_1}{dt}=-\vec S_1\times\vec B_1-J(\vec S_1\times\vec S_2),
\end{equation}
\begin{equation}
	\frac{d\vec S_2}{dt}=-\vec S_2\times\vec B_2+J(\vec S_1\times\vec S_2),
\end{equation}
\begin{equation}
	\frac{d}{dt}(\vec S_1\cdot\vec S_2)=(\vec S_1\times\vec S_2)\cdot(\vec B_1-\vec B_2),
\end{equation}
\begin{widetext}
\begin{equation}
	\frac{d}{dt}(\vec S_1\times\vec S_2)=-\frac{1}{2}(\vec S_1\times\vec S_2)\times(\vec B_1+\vec B_2)+\left(\vec Q-\frac{2}{3}(\vec S_1\cdot\vec S_2)\right)\cdot(\vec B_1-\vec B_2)+\frac{J}{2}(\vec S_1-\vec S_2),
\end{equation}
\begin{multline}
	\frac{d}{dt}\vec Q_{ij}=-\frac{1}{2}\left(\frac{1}{2}(\vec S_1\times\vec S_2)_i(\vec B_1-\vec B_2)_j+\frac{1}{2}(\vec S_1\times\vec S_2)_j(\vec B_1-\vec B_2)_i-\frac{1}{3}(\vec S_1\times\vec S_2)\cdot(\vec B_1-\vec B_2)\delta_{ij}\right)\\
	-\frac{1}{2}\varepsilon_{ilm}\vec Q_{lj}(\vec B_1+\vec B_2)_m-\frac{1}{2}\varepsilon_{jlm}\vec Q_{li}(\vec B_1+\vec B_2)_m.
\end{multline}
\end{widetext}
It is interesting to note that some terms contain the sum of the two magnetic field while other couple to the difference. In consequence, for the quantum-dot system under investigation, some terms are only sensitive to the exchange field as the external field acts on both spins in the same way, while other terms are sensitive to both field contributions.

%\bibliography{/home/bjoern/LaTeX/Bibtex/Meine_Bibliothek}

%Merlin.mbs v4.21 2009-07-09.
%

\end{document}